\documentclass[a4paper,11pt]{article}
\usepackage[dvips]{graphicx}
\begin{document}

\author{A. Raiteri}
\title{A realistic interpretation of the density matrix\\
II: The non-relativistic case}
\date{July, 2000}
\maketitle

\section{Introduction}\label{sec_1}
In a recent paper \cite{ar} we proposed a realistic interpretation of the
Schr\"odinger equation for density matrices, in which the difference between
the position arguments of the density matrix is considered as an objective
extra space dimension. In the case of a free particle, where the potential
$V\left(x\right)$ vanishes, we found solutions which are perfectly localized
both in position space and in momentum space; these solutions behave exactly
as non-relativistic point-like particles moving at constant speed, with the
correct values for all observable quantities. In the general case, where
$V\left(x\right)\ne 0$, we pointed out that the natural frequencies of the
stationary states in the density matrix representation correspond to the
difference between two energy levels in the original quantum system; since
the ``jumps'' between energy levels are observable, while the individual
energy levels are not, we trivially deduced that the observable natural
frequencies are the same in both representations.

In the first paper the non-relativistic case was treated mainly as an
introduction to the relativistic case; in this second paper we will study it
in more detail, examining the correspondence between our new representation
and the standard representation of non-relativistic quantum mechanics, based
on the Schr\"odinger equation for pure states. In Section \ref{sec_2} we
will first summarize the main results of the previous paper, and then we
will give a qualitative definition of a quantum trajectory (describing an
individual physical system), as opposed to a statistical mixture (describing
an ensemble of physical systems); as a consequence, we will reject two basic
principles of standard quantum mechanics, namely the superposition principle
and the wave function collapse. In Section \ref{sec_3} we will show the
results obtained by applying our approach to simple potentials: by examining
the linear harmonic oscillator we will shed a new light on the supposed
equivalence between energy eigenvectors and stationary states; besides,
the delta barrier potential will provide a semi-classical explanation of the
tunnel effect, in which an essential feature is the introduction of internal
degrees of freedom even for spinless particles. Finally in Section \ref{sec_4}
we will present our conclusions.

\section{The density matrix representation and the superposition principle}
\label{sec_2}
The standard quantum description of a non-relativistic particle moving in
one space dimension is based on a wave function $\psi\left(x\right)$, whose
time evolution is given by the Schr\"odinger equation:
\begin{equation}\label{eq_2_1}
i\hbar\frac{\partial\psi}{\partial t}=\textrm{H}\,\psi=-\frac{\hbar^2}{2m}
\frac{\partial^2\psi}{\partial x^2}+V\!\left(x\right)\psi
\end{equation}
We will refer to equation (\ref{eq_2_1}) as to the Schr\"odinger equation
for pure states. The time evolution generated by (\ref{eq_2_1}) belongs to
the group of unitary transformations; a new representation of this group,
the density matrix representation, is obtained by means of the fundamental
relation
\begin{equation}\label{eq_2_2}
\varphi\left(x,y\right)=\psi\left(x\right)\psi^*\left(y\right)
\end{equation}
If we apply an arbitrary unitary transformation to the pure state $\psi
\left(x\right)$, the relation (\ref{eq_2_2}) enables us to obtain the
same transformation in the density matrix representation. Specifically, if
we consider the hamiltonian operator H, the momentum operator P and the
position operator Q as generators respectively of time translations, space
translations and momentum translations, we easily obtain the following
expressions for the same generators in the new representation:
\begin{eqnarray}
\textrm{H}\,\varphi&=&-\frac{\hbar^2}{2m}\left(\frac{\partial^2\varphi}
{\partial x^2}-\frac{\partial^2\varphi}{\partial y^2}\right)+\left[V\!
\left(x\right)-V\!\left(y\right)\right]\varphi\nonumber\\[6pt]
\textrm{Q}\,\varphi&=&\left(x-y\right)\varphi\label{eq_2_3}\\[6pt]
\textrm{P}\varphi&=&-i\hbar\left(\frac{\partial\varphi}{\partial x}
+\frac{\partial\varphi}{\partial y}\right)\nonumber
\end{eqnarray}
The time evolution in the density matrix representation is then simply
\begin{equation}   \label{eq_2_4}
i\hbar\frac{\partial\varphi}{\partial t}=
-\frac{\hbar^2}{2m}\left(\frac{\partial^2\varphi}{\partial x^2}-
\frac{\partial^2\varphi}{\partial y^2}\right)+\left[V\!\left(x\right)-
V\!\left(y\right)\right]\varphi
\end{equation}
In the usual interpretation, equation (\ref{eq_2_4}) describes the time
evolution of mixed states, and therefore its solutions have only a
statistical meaning: they are simply useful tools for computing probability
amplitudes and expectation values. On the contrary, in our approach the
solutions of equation (\ref{eq_2_4}) are considered as real objective fields
describing individual physical systems. To avoid misunderstandings, in the
rest of this paper we will use the expression ``density matrix'' when we
mean the usual statistical interpretation and we will use the new expression
``quantum matrix'' when we mean a real objective field.

A quantum matrix $\varphi\left(x,y\right)$ depends upon two position
coordinates $x$ and $y$; if we want to consider $\varphi\left(x,y\right)$
as a real objective field then we must introduce an objective extra space
dimension. Therefore, we will define a new pair of position coordinates
\begin{equation}\label{eq_2_5}
x_S=\frac{1}{2}\left(x+y\right)\qquad\qquad\qquad x_D=x-y
\end{equation}
and we will interpret $x_S$ as the ``physical'' position coordinate, while
$x_D$ will be an ``auxiliary'' position coordinate, having observable effects
only around the point $x_D=0$; both $x_S$ and $x_D$, however, will be
considered as objective position coordinates (note that in the definition of
$x_D$ we have inverted the sign convention with respect to our first paper
\cite{ar}). In the rest of this paper we will write loosely $\varphi\left(x,y
\right)$ or $\varphi\left(x_S,x_D\right)$, meaning the same field expressed
in two different coordinate systems.

The transition from position space to momentum space is obtained by means
of the two-dimensional Fourier transform:
\begin{equation} \label{eq_2_6}
\Phi\left(k_x,k_y\right)=\frac{1}{2\pi}\int\!\int{\varphi\left(x,y\right)
e^{-i k_x\!x+i k_y\!y}\mathrm{d}x\,\mathrm{d}y}
\end{equation}
Again we define a new pair of momentum coordinates:
\begin{equation}\label{eq_2_7}
k_S=\frac{1}{2}\left(k_x+k_y\right)\qquad\qquad\qquad k_D=k_x-k_y
\end{equation}
and again $k_S$ will be the ``physical'' momentum coordinate, while $k_D$
will be an ``auxiliary'' momentum coordinate, having observable effects only
around the point $k_D=0$.

Since we have two position coordinates and two momentum coordinates, we
naturally can define two position operators and two momentum operators. The
operators already defined in (\ref{eq_2_3}) may be rewritten as:
\begin{equation}\label{eq_2_8}
\textrm{Q}_D\,\varphi=x_D\,\varphi\qquad\qquad\qquad\quad
\textrm{P}\!_D\,\varphi=-i\hbar\frac{\partial\varphi}{\partial x_S}\quad
\end{equation}
In addition, we introduce the new operators $\textrm{Q}_S$ and $\textrm{P}
\!_S$ by means of the following definition:
\begin{equation}\label{eq_2_9}
\textrm{Q}_S\,\varphi=x_S\,\varphi\qquad\qquad\qquad\quad
\textrm{P}\!_S\,\varphi=-i\hbar\frac{\partial\varphi}{\partial x_D}\quad
\end{equation}
In momentum space the operators $\textrm{P}\!_D$ and
$\textrm{P}\!_S$ are simply given by:
\begin{equation}\label{eq_2_10}
\textrm{P}\!_D\,\Phi=\hbar k_D\,\Phi\qquad\qquad\qquad\quad
\textrm{P}\!_S\,\Phi=\hbar k_S\,\Phi\quad
\end{equation}
It is very important here to note that $\textrm{Q}_S$ and $\textrm{P}\!_S$
are commuting operators, i.e. $\left[\textrm{Q}_S,\textrm{P}\!_S\right]=0$;
this property is fundamental for our approach, because it means that they
have common eigenvectors. These eigenvectors have the form
\begin{equation}\label{eq_2_11}
\varphi=\delta\left(x_S-x_0\right)e^{i k_0 x_D}
\end{equation}
or, in momentum space,
\begin{equation}\label{eq_2_12}
\Phi=\delta\left(k_S-k_0\right)e^{-i k_D x_0}
\end{equation}
Since $x_S$ and $k_S$ are the ``physical'' position and momentum
coordinates, it is clear that the field defined in (\ref{eq_2_11}) and
(\ref{eq_2_12}) is perfectly localized in position space at $x_S=x_0$ and in
momentum space at $p_S=p_0=\hbar k_0$. Therefore in the density matrix
representation the Heisenberg uncertainty principle is not true and we
conclude, as in our previous paper, that the Heisenberg principle has no
ontological meaning; rather, it is just a shortcoming of the standard
quantum formalism based on pure states. A similar point of view is expressed
by Olavo \cite{ola}.

While the two operators $\textrm{Q}_D$ and $\textrm{P}\!_D$ are interesting
just because they generate translations respectively in momentum space and
in position space, the two operators $\textrm{Q}_S$ and $\textrm{P}\!_S$
are more strictly related to the measurement of the corresponding physical
quantities, as we will now see. To define the observable quantities in the
density matrix representation, we start from the expressions for the mean
values in the original representation and then exploit the fundamental
relation (\ref{eq_2_2}); we obtain:
\begin{eqnarray}
Q&=&\int{\left(\textrm{Q}_S\,\varphi\right)\Big|_{x_D=0}\textrm{d}x_S}
\label{eq_2_13}\\[12pt]
P&=&\int{\left(\textrm{P}\!_S\,\varphi\right)\Big|_{x_D=0}\textrm{d}x_S}
\label{eq_2_14}\\[8pt]
E&=&\int{\left[\frac{1}{2m}\,\textrm{P}_{\!S}^2+V\!\left(\textrm{Q}_S\right)
\right]\!\varphi\,\Bigg|_{x_D=0}\textrm{d}x_S}
\label{eq_2_15}
\end{eqnarray}
The above definitions confirm our interpretation of $x_D$ as an ``auxiliary''
position coordinate, having observable effects only around $x_D=0$.
The same quantities may be written in momentum space, obtaining analogous
expressions; for instance, the momentum $P$ may be written as
\begin{equation}\label{eq_2_16}
P=\int{\left(\textrm{P}\!_S\,\Phi\right)\Big|_{k_D=0}\textrm{d}k_S}
\end{equation}
Since we do not assign a statistical meaning to the quantum matrix
$\varphi$, we will not interpret the expressions
(\ref{eq_2_13})-(\ref{eq_2_15}) as mean values. On the contrary, we will
consider (\ref{eq_2_13}) as the center of mass of the system, while
(\ref{eq_2_14}) and (\ref{eq_2_15}) will be respectively the total
momentum and the total energy.

If we Fourier transform the quantum matrix $\varphi$ with respect to $x_D$
alone, we obtain the well known Moyal-Wigner transformation:
\begin{equation} \label{eq_2_17}
F\left(x_S,p_S\right)=\frac{1}{2\pi\hbar}\int{\varphi\left(x_S,x_D\right)
e^{-i\frac{\scriptstyle p_S}{\scriptstyle\hbar}x_D}\mathrm{d}x_D}
\end{equation}
first introduced by Wigner \cite{wig} in 1932.
The Wigner function $F\left(x_S,p_S\right)$ has some valuable properties:
the common eigenvectors of the operators $\textrm {Q}_S$ and $\textrm{P}\!
_S$ are represented simply by products of delta functions, i.e. $F\left(
x_S,p_S\right)=\delta\left(x_S-x_0\right)\delta\left(p_S-p_0\right)$;
besides, those operators which depend only on $\textrm{Q}_S$ and
$\textrm{P}\!_S$ become c-numbers in the Wigner representation; finally,
if $G$ is an observable quantity depending on position and momentum through
the function $g\left(x,p\right)=g_1\left(x\right)+g_2\left(p\right)$, then
in the Wigner representation we can write the expression
\begin{equation} \label{eq_2_18}
G=\int\!\int{g\left(x_S,p_S\right) F\left(x_S,p_S\right)\mathrm{d}x_S\,
\mathrm{d}p_S}
\end{equation}
i.e. $F\left(x_S,p_S\right)$ weighs the function $g\left(x_S,p_S\right)$
over the whole phase space $(x_S,p_S)$; the application of (\ref{eq_2_18})
to the quantities defined in (\ref{eq_2_13})-(\ref{eq_2_15}) is
straightforward. Thus the Wigner function is a useful tool for establishing
relations between the quantum description and the classical description
based on phase space; indeed, it provides the basis for Moyal's deformation
quantization \cite{moy}, which is an autonomous formulation of quantum
mechanics, alternative to the more familiar Hilbert space and path integral
quantizations.

However, in the usual statistical interpretation of the density matrix,
expression (\ref{eq_2_18}) yields the mean value of the quantity $G$, and
therefore the Wigner function seems to play the role of a joint probability
distribution in phase space. But even if the function $\varphi\left(x,y
\right)$ satisfies all the properties required for being a well defined
density matrix, it is in any case possible for the Wigner function to be
negative in some phase space region. This is the well known ``negativity
problem'', which prevents a complete analogy between the classical
description in phase space and the quantum description in the Wigner
representation. On the contrary, if we consider the quantum matrix
$\varphi\left(x,y\right)$ as a real objective field, describing the state
of an individual physical system, we do not assign to the Wigner function
any statistical meaning, and therefore it may safely have negative values:
in our interpretation, the ``negativity problem'' is no problem at all.

Now we want to study the difference between the time evolution of a quantum
system in the density matrix representation and the time evolution of the
corresponding classical system in phase space. We will use the Wigner
representation, but the results may be easily carried to position space
and momentum space, since we will express them in operator form. The
quantum evolution, obtained from (\ref{eq_2_4}), is simply given by:
\begin{equation}   \label{eq_2_19}
i\hbar\frac{\partial F_Q}{\partial t}=H_Q F_Q=
\left[\,\frac{1}{m}\,\textrm{P}\!_S\,\textrm{P}\!_D+
V\!\left(\!\textrm{Q}_S\!+\!\frac{1}{2}\textrm{Q}_D\!\right)-
V\!\left(\!\textrm{Q}_S\!-\!\frac{1}{2}\textrm{Q}_D\!\right)
\right]F_Q
\end{equation}
To obtain the corresponding classical evolution we start from the continuity
equation for the classical joint probability density in phase space:
\begin{equation}   \label{eq_2_20}
\frac{\partial F_C}{\partial t}+\frac{\partial F_C}{\partial x}\,
\frac{p}{m}-\frac{\partial F_C}{\partial p}\,V'=0
\end{equation}
where $V'$ is the space derivative of the potential $V$. Translating
(\ref{eq_2_20}) to operator form we then obtain
\begin{equation}   \label{eq_2_21}
i\hbar\frac{\partial F_C}{\partial t}=H_C F_C=
\left[\,\frac{1}{m}\,\textrm{P}\!_S\,\textrm{P}\!_D+
V'\left(\textrm{Q}_S\right)\textrm{Q}_D\right]F_C
\end{equation}
Expressions (\ref{eq_2_19}) and (\ref{eq_2_21}) are quite similar; indeed,
if we expand $V\left(x\right)$ in powers of $x$, we can write:
\begin{equation}   \label{eq_2_22}
H_Q-H_C=\sum_{n=1}^\infty\frac{1}{\left(2n+1\right)!\,2^{2n}}V^{(2n+1)}
\left(\textrm{Q}_S\right)\textrm{Q}_D^{(2n+1)}
\end{equation}
The difference between the quantum evolution and the classical evolution
involves the space derivatives $V^{(2n+1)}$ with $n\ge 1$. Therefore the
quantum evolution and the classical evolution are equivalent if $V=0$ (free
particle), $V=kx$ (particle moving in a uniform force field) and $V=kx^2$
(linear harmonic oscillator). This is a rather surprising result since we
know that the quantum energy spectrum of the linear harmonic oscillator
is discrete, and this is a highly non-classical feature; we will examine
this point in detail in the next section.

Now we turn to the problem of finding localized solutions in the density
matrix representation; the existence of such solutions is fundamental for
our approach. It is well known that to each density matrix satisfying the
Schr\"odinger equation (\ref{eq_2_4}) we can associate a quasi-classical
trajectory, namely the time evolution of the mean values $\langle x\left(t
\right)\rangle$ and $\langle p\left(t\right)\rangle$ of position and momentum;
this is of course a consequence of the Ehrenfest theorem:
\begin{equation}   \label{eq_2_23}
\dot{\langle x\rangle}=\frac{1}{m}\,\langle p\rangle\qquad\qquad\qquad\qquad
\dot{\langle p\rangle}=-\langle V'\!\left(x\right)\rangle\qquad\qquad
\end{equation}
From the trivial relation
\begin{equation}   \label{eq_2_24}
\langle V'\!\left(x\right)\rangle=V'\!\left(\langle x\rangle\right)+
\sum_{n=2}^\infty\frac{1}{n!}\,V^{\left(n+1\right)}\!\left(\langle
x\rangle\right)\,\langle\left(x-\langle x\rangle\right)^n\rangle
\end{equation}
it then follows that the equations (\ref{eq_2_23}) define a classical
trajectory only if the derivatives $V^{\left(n\right)}$ vanish for $n\!>\!2$
(again the free particle and the linear harmonic oscillator), while in
the general case the trajectory will be quasi-classical if the moments
$\langle\left(x-\langle x\rangle\right)^n\rangle$ are small, i.e.\ if the
wave-packet is well localized in position space around its mean value
$\langle x\left(t\right)\rangle$.

In standard quantum mechanics, due to the Heisenberg uncertainty principle,
wave-packets normally spread out with time; only in some special cases this
does not happen, the most famous example being the coherent states of the
linear harmonic oscillator first introduced by Schr\"odinger in 1926
\cite{schr}. However, in the density matrix representation the Heisenberg
uncertainty principle is not true, and therefore it should be always possible
to find solutions which do not spread out with time and remain well localized
around their center of mass $\langle x\left(t\right)\rangle$, in the sense
that their ``matter density'' $\varphi\left(x_S,x_D,t\right)\Big|_{x_D=0}$ is
significantly different from zero only in a small region around $x_S=\langle
x\left(t\right)\rangle$. In our approach, the existence of such solutions is
of the utmost importance and allows us to state the following ``localization
postulate'': the only physical solutions of the Schr\"odinger equation
(\ref{eq_2_4}) are those which remain well localized around their center
of mass $\langle x\left(t\right)\rangle$ in the limit $t\to\pm\infty$;
by ``physical'' we mean ``representing individual physical systems'', as
opposed to statistical mixtures or ensembles. In the rest of this paper,
we will use the definition ``quantum trajectories'' to label such solutions:
thus a quantum trajectory is the time evolution of a quantum matrix, as
defined at the beginning of the present section.

Clearly, our localization postulate is expressed as a qualitative statement,
since the concept of ``well localized'' wave packet is not sharply defined.
However, even in its rudimentary form, this postulate implies an important
difference between our approach and the standard interpretation of quantum
mechanics: in standard QM, immediately before the measure of an observable
$O$ (for instance the position), the state of the system is usually a linear
superposition of eigenstates of the associated hermitian operator; then,
when the measure is performed and the result $O_n$ is obtained, the state
suddenly collapses to the eigenvector $|O_n\rangle$. This wave function
collapse has always been an obscure feature of the standard QM
interpretation: many alternative explanations have been proposed (many
worlds splitting \cite{mwi}, environment induced decoherence \cite{deco1,
deco2}, spontaneous localization \cite{grw,pearl}, \dots) but until now none
of these alternative explanations seems to be universally accepted by the
academic community.

On the contrary, in our approach the physical wave-packets are always well
localized around their center of mass $\langle x\left(t\right)\rangle$ and
therefore no wave function collapse is needed. When a measure of position is
performed, and the result is a certain position $x_0$, this means that the
wave-packet immediately before the measure was already well localized around
the point $x_0$: our localization postulate explicitly prevents the spreading
of the physical wave function $\varphi\left(x,y,t\right)$ over a wide space
region as $t\to\infty$.

Besides eliminating the need of the wave function
collapse, our localization postulate implicitly negates the validity of the
superposition principle: in our approach, a linear combination of physical
solutions is not a physical solution any more, even if it satisfies the linear
motion equation (\ref{eq_2_4}). This is a trivial consequence of the fact that
in general an arbitrary superposition of localized wave-packets is not a
localized wave-packet.

However, a linear superposition of quantum trajectories may have a statistical
interpretation. Let's consider the set $\left\{\varphi_\lambda\right\}$
of all quantum trajectories, where $\lambda$ is some appropriate index (for
instance, $\lambda$ may be the center of mass and momentum at time $t=0$,
together with some non-classical internal state). Then the integral
\begin{equation}\label{eq_2_25}
\varphi\left(x,y,t\right)=\int{f\left(\lambda\right)\,
\varphi_\lambda\left(x,y,t\right)\textrm{d}\lambda}
\end{equation}
where $f\left(\lambda\right)$ is a non-negative probability distribution,
represents a statistical ensemble of particles; in our approach, the solutions
defined by (\ref{eq_2_25}) play the same role as the density matrices in
standard quantum mechanics. Besides, our approach rejects those solutions of
the Schr\"odinger equation (\ref{eq_2_4}) which are neither quantum
trajectories nor positive superpositions of quantum trajectories: since they
do not represent neither individual particles nor statistical ensembles of
particles, we conclude that they are just mathematical objects with no
physical meaning.

If we now apply the Moyal-Wigner transformation to the density matrix
(\ref{eq_2_25}), we obtain the Wigner function
\begin{equation}\label{eq_2_26}
F\left(x_S,p_S,t\right)=\int{f\left(\lambda\right)\,
F_\lambda\left(x_S,p_S,t\right)\textrm{d}\lambda}
\end{equation}
where $F_\lambda\left(x_S,p_S,t\right)$ are the individual Wigner functions
associated to the quantum trajectories $\varphi_\lambda\left(x,y,t\right)$.
Note that both the individual functions $F_\lambda$ and the average function
$F$ may safely have negative values, since neither represents a probability
density. The only probability density in (\ref{eq_2_26}) is the function
$f\left(\lambda\right)$, which is non-negative by definition.

Let's now examine the problem of finding the quantum trajectories for a
specified potential $V\left(x\right)$. We will suppose that $V\left(x\right)$
vanishes as $x\to\pm\infty$: in classical terms, this means that the particle
is subject to a force only in a limited space region, while outside this
region it moves freely at constant speed. The first step consists in writing
the Schr\"odinger equation (\ref{eq_2_4}) in the form
\begin{equation}\label{eq_2_27}
\frac{\partial^2}{\partial x_S\partial x_D}\varphi=\frac m{\hbar^2}\left[
V\left(x_S+\frac 1 2 x_D\right)-V\left(x_S-\frac 1 2 x_D\right)-i\hbar
\frac\partial{\partial t}\right]\varphi
\end{equation}
From (\ref{eq_2_27}) it is easily seen that, given two arbitrary functions
$f\left(x_S,t\right)$ and $g\left(x_D,t\right)$, we may formally impose the
boundary conditions
\begin{equation}\label{eq_2_28}
\varphi\left(x_S,x_D,t\right)\Big|_{x_D=0}=f\left(x_S,t\right)
\quad\textrm{and}\quad
\frac{\partial\varphi}{\partial x_D}\left(x_S,x_D,t\right)\Big|_{x_S=0}=
g\left(x_D,t\right)\quad
\end{equation}
and then we may extend (\ref{eq_2_28}) to a complete solution $\varphi\left(
x_S,x_D,t\right)$ by means of the Taylor expansion
\begin{equation}\label{eq_2_29}
\varphi\left(x_S,x_D,t\right)=f\left(x_S,t\right)+\sum_{n=1}^\infty
\frac1{n!}\,\frac{\partial^n\varphi}{\partial x_D^n}\bigg|_{x_D=0}x_D^n
\end{equation}
where the derivatives are obtained by differentiating $n$ times
(\ref{eq_2_27}) with respect to $x_D$ and then imposing $x_D=0$.

At first sight, this result is rather paradoxical: it implies that we may
choose an arbitrary time evolution for the center of mass of our system and
then obtain a solution of the Schr\"odinger equation which satisfies this
time evolution! However, there are at least two reasons why the above
procedure does not always produce acceptable results: the first reason
follows from the fact that an arbitrary solution of the Schr\"odinger equation
(\ref{eq_2_4}) must be a linear combination of the form
\begin{equation}\label{eq_2_30}
\varphi\left(x,y,t\right)=\int\!\!\textrm{d}k_1\int\!\!\textrm{d}k_2\ C\!
\left(k_1,k_2\right)f_{k1}\left(x\right)\,f_{k2}^{\,*}\left(y\right)
e^{-\frac{\scriptstyle i}{\scriptstyle\hbar}\,\left(E_{k1}-E_{k2}\right)\,t}
\end{equation}
where $f_k\left(x\right)$ are the energy eigenvectors of the original
Schr\"odinger equation (\ref{eq_2_1}) with associated eigenvalues $E_k$; of
course, if the energy spectrum is discrete then the integrals in
(\ref{eq_2_30}) must be replaced by summations. Therefore the function
$f\left(x_S,t\right)$ in (\ref{eq_2_28}) cannot be chosen arbitrarily, since
it must be equal to (\ref{eq_2_30}), in the case $x=y=x_S$, for some choice of
the coefficients $C\!\left(k_1,k_2\right)$. For instance, we know that for
the linear harmonic oscillator the solutions must be periodic in time, i.e.\
$\varphi\left(x,y,t\right)=\varphi\left(x,y,t+T_0\right)$ where $T_0$ is the
revolution time of the oscillator; if we choose a function $f\left(x_S,t
\right)$ which is not periodic, then it will be impossible to obtain from it
a complete solution $\varphi\left(x_S,x_D,t\right)$ by means of
(\ref{eq_2_28}), i.e.\ the Taylor series (\ref{eq_2_29}) will diverge.

The second reason is that the solutions obtained from (\ref{eq_2_27}) and
(\ref{eq_2_28}), even if they are well defined from a mathematical point of
view, will in general violate some fundamental physical principle, such as the
Ehrenfest theorem (\ref{eq_2_23}), the energy conservation principle or the
unitarity condition
\begin{equation}\label{eq_2_31}
\frac\partial{\partial t}\int\varphi\left(x_S,x_D,t\right)\Big|_{x_D=0}
\textrm{d}x_S=0
\end{equation}
which may be regarded as an expression of the non-relativistic principle of
mass conservation. Therefore, our next step consists in stating clearly the
assumptions under which the above cited principles may be derived from the
Schr\"odinger equation; we will then require that our quantum trajectories
satisfy these assumptions. The unitarity property (\ref{eq_2_31}) may be
derived from the following condition:
\begin{equation}\label{eq_2_32}
\frac{\partial\varphi}{\partial x_D}\bigg|_{x_D=0}\to\,0\qquad\textrm{as}
\qquad x_S\to\pm\infty
\end{equation}
A sufficient condition to obtain the Ehrenfest theorem (\ref{eq_2_23}) is
\begin{equation}\label{eq_2_33}
x_S\frac{\partial\varphi}{\partial x_D}\bigg|_{x_D=0}\to\,0\quad\textrm{and}
\quad\frac{\partial^2\varphi}{\partial x_D^2}\bigg|_{x_D=0}\to\,0\qquad
\textrm{as}\qquad x_S\to\pm\infty
\end{equation}
and finally the energy conservation principle follows from
\begin{equation}\label{eq_2_34}
\frac{\partial^3\varphi}{\partial x_D^3}\bigg|_{x_D=0}\to\,0\qquad\textrm{as}
\qquad x_S\to\pm\infty\quad
\end{equation}

Thus the assumptions (\ref{eq_2_32})-(\ref{eq_2_34}) depend only on the field
$\varphi$ and its three first derivatives with respect to $x_D$ computed at
$x_D=0$. Besides, from the expressions (\ref{eq_2_13})-(\ref{eq_2_15}) we see
that the same is true for the three observable quantities, i.e.\ center of 
mass, momentum and energy (in this case only the first two derivatives are
involved). If we now set
\begin{equation}\label{eq_2_35}
\varphi^{\left(0\right)}=\varphi\big|_{x_D=0}\qquad\quad\textrm{and}
\qquad\quad\varphi^{\left(n\right)}=
\frac{\partial^n\varphi}{\partial x_D^n}\bigg|_{x_D=0}\quad n=1,2,3\quad
\end{equation}
where $\varphi^{\left(0\right)}$ and $\varphi^{\left(n\right)}$ depend only
on $x_S$ and $t$, we easily derive from the Schr\"odinger equation 
(\ref{eq_2_27}) the following relations
\begin{eqnarray}
\frac{\partial\varphi^{\left(1\right)}}{\partial x_S}&=&-i\frac m\hbar
\frac{\partial\varphi^{\left(0\right)}}{\partial t}\label{eq_2_36}\\[4pt]
\frac{\partial\varphi^{\left(2\right)}}{\partial x_S}&=&-i\frac m\hbar
\frac{\partial\varphi^{\left(1\right)}}{\partial t}+\frac m{\hbar^2}
V'\left(x_S\right)\varphi^{\left(0\right)}\label{eq_2_37}\\[4pt]
\frac{\partial\varphi^{\left(3\right)}}{\partial x_S}&=&-i\frac m\hbar
\frac{\partial\varphi^{\left(2\right)}}{\partial t}+2\frac m{\hbar^2}
V'\left(x_S\right)\varphi^{\left(1\right)}\label{eq_2_38}
\end{eqnarray}
From the above considerations we deduce that we do not need to know the
complete quantum trajectory $\varphi\left(x_S,x_D,t\right)$: all properties
of physical interest follow from the knowledge of the four functions
$\varphi^{\left(i\right)}\left(x_S,t\right)$, with $i=0,\dots,3$.
We only need to impose that these four functions satisfy the relations
(\ref{eq_2_36})-(\ref{eq_2_38}) together with the boundary conditions
(\ref{eq_2_32})-(\ref{eq_2_34}), and that $\varphi^{\left(0\right)}\left(
x_S,t\right)$ be a linear combination as in (\ref{eq_2_30}), with $x=y=x_S$,
for some choice of the coefficients $C\!\left(k_1,k_2\right)$. Surprisingly
enough, the relations (\ref{eq_2_36})-(\ref{eq_2_38}), obtained from the
quantum evolution (\ref{eq_2_19}), are exactly the same that we would obtain
from the classical evolution (\ref{eq_2_21}). Therefore, if a classical
trajectory $\varphi^{\left(0\right)}\left(x_S,t\right)=\delta\left(x_S-
x_{cl}\left(t\right)\right)$ may be expressed as a linear combination
(\ref{eq_2_30}), then it can be exactly reproduced in the quantum domain:
this is more likely to happen in the case of open trajectories, for which the
quantum energy spectrum is continuous, rather than for closed trajectories,
where the quantum energy spectrum is discrete and most classical energy levels
are forbidden; let's examine these two cases in more detail.

An open trajectory represents a particle which in the limit $t\to\pm\infty$
behaves like a free particle; at some finite time, the particle interacts with
the potential $V\left(x\right)$ and undergoes a scattering process. In one
space dimension, there are only two possible outgoing directions, i.e.\ the
particle may be transmitted or reflected; this is true both in the classical
and in the quantum description. However, in the quantum domain we must take
into account a new, highly non-classical feature, namely the tunnel effect:
particles with energy less than the potential peak $V_{max}$ may cross
the barrier, and viceversa particles with energy greater than the barrier
peak may be reflected. Therefore, in the quantum case we have to remove the
classical restriction that all particles with kinetic energy $m v^2/2>V_{max}$
are transmitted while all particles with $m v^2/2< V_{max}$ are reflected.
Thus, knowing the energy of a quantum particle is not enough to determine
whether it will be transmitted or reflected; since we believe that the physical
laws are deterministic, we deduce that the property of an incoming wave-packet
to be transmitted or reflected depends also on the form of the wave-packet, not
only on its energy: in other words, besides the center of mass and momentum,
the state of a quantum particle must include also some internal, or hidden,
degree of freedom even in the case of a spinless uncharged particle.

Furthermore, the existence of non-classical tunnelling particles implies
that their wave-packet must lose its sharp localization while crossing the
barrier region; if this were not true, then (\ref{eq_2_23}) and
(\ref{eq_2_24}) would yield a quasi-classical trajectory for the center
of mass $\langle x\left(t\right)\rangle$, thus preventing the appearance of
non-classical effects. This is the reason why in our localization postulate
we explicitly introduced the limit $t\to\pm\infty$: in general, an open
quantum trajectory will be well localized when $t\to -\infty$, i.e.\ when
it is still far away from the potential region; then, while interacting
with the external potential, it will spread out to some extent; finally, as
$t\to +\infty$ and the particle leaves the potential region, it will recover
its localization. In the next section, we will see a practical application of
our approach to the tunnel effect.

Let's now consider briefly the case of closed trajectories. It is natural
to assume that quantum closed trajectories are periodic, i.e.\
$\varphi^Q_{cl}\left(x,y,t+T_{cl}\right)=\varphi^Q_{cl}\left(x,y,t\right)$
where $T_{cl}$ is the revolution time. The following condition must then be
fulfilled
\begin{equation}\label{eq_2_39}
\frac{1}{\hbar}\left(E_{k1}-E_{k2}\right)=n\frac{2\pi}{T_{cl}}
\end{equation}
for some integer number $n$, where again $E_{k1}$, $E_{k2}$ are energy
eigenvalues of the Schr\"odinger equation for pure states. Since normally
the revolution time $T_{cl}$ depends on the energy of the particle,
(\ref{eq_2_39}) may then be interpreted as a quantization condition for the
energy spectrum in the quantum domain; an important exception is provided
by the linear harmonic oscillator, where notoriously the revolution
time is independent from the energy. In our approach the condition
(\ref{eq_2_39}) is responsible for the existence of discrete energy spectra.

\section{Applications}\label{sec_3}
\subsection{Free particle}\label{sec_3_1}
In the free particle case, the quantum trajectories are simply given by
\begin{equation}\label{eq_3_1_1}
\varphi_Q\left(x,y,t\right)=\delta\Big(x_S-x_0-v_0\,t\Big)
e^{\frac{\scriptstyle i}{\scriptstyle\hbar}\,m\,v_0 x_D}
\end{equation}
that is, the quantum description is equivalent to the classical description.
We will briefly show, by means of the Moyal-Wigner transformation, the
classical probability densities associated to some well known quantum
states.

In the case of a plane wave we have
\begin{equation}\label{eq_3_1_2}
\psi\left(x,t\right)=e^{i\,\frac{p_0}{\hbar}\,\left(x-\frac{p_0}{2m}\,t
\right)}\qquad\Longrightarrow\qquad F\left(x,p,t\right)=\delta\left(p-p_0
\right)\quad
\end{equation}
Therefore, the plane wave describes an ensemble of classical particles
moving at speed $v_0=\frac{1}{m}\,p_0$ uniformly distributed over the $x$
axis (the normalization is one particle per unit of length).

In the case of a particle perfectly localized at $x=0$, whose initial
quantum state is given by $\psi\left(x,0\right)=2\pi\hbar\,\delta\left(x
\right)$, we have:
\begin{equation}\label{eq_3_1_3}
\psi\left(x,t\right)=\int{\!e^{i\,\frac{p}{\hbar}\,\left(x-\frac{p}{2m}
\,t\right)}}\,\textrm{d}p\qquad\Longrightarrow\qquad F\left(x,p,t\right)=
\delta\left(x-\frac{p}{m}\,t\right)\quad
\end{equation}
Thus we obtain an ensemble of classical particles following the trajectories
\mbox{$x=v t$}, with $v=\frac{1}{m}\,p$, uniformly distributed in momentum
space (the normalization is one particle per unit of momentum).

Finally, for a gaussian state we have:
\begin{equation}\label{eq_3_1_4}
\psi\!\left(x,t\right)\!=\!\!\sqrt\frac{1}{\sqrt{2\pi}\Delta x_0\,\alpha\!\left(
t\right)}\,e^{\!-\frac{\scriptstyle x^2}{\scriptstyle 4 {\Delta x_0\!\!}^2\alpha
\left(t\right)}}\Longrightarrow F\!\left(\!x,p,t\right)\!=\!\!
\frac{1}{\pi\!\hbar}\,e^{\!-2{\Delta x_0\!\!}^2\frac{\scriptstyle p^2}
{\scriptstyle \hbar^2}-\frac{\scriptstyle \left(x-{p\!/\!m}\,t\right)^2}
{\scriptstyle 2{\Delta x_0\!\!}^2}}
\end{equation}
where $\alpha\left(t\right)\!=\!1\!+\!\frac{i\hbar}{2m{\Delta x_0\!}^2}\,t$.
The marginal deviations for the classical joint distribution
$F\left(x,p,t\right)$ at $t\!=\!0$ are $\Delta x_0$ and
$\Delta p_0\!=\!\frac{\hbar}{2\Delta x_0}$, yielding the minimum deviation
product allowed by the Heisenberg uncertainty relations. For $t\!>\!0$
$\Delta p$ remains constant, while $\Delta x$ grows according to the law
\mbox{$\Delta x=\sqrt{{\Delta x_0\!}^2+\frac{\hbar^2}{4m^2{\Delta x_0\!}^2}
\,t^2}$}, in agreement with the quantum predictions.

\subsection{The linear harmonic oscillator}\label{sec_3_2}
Let's consider the case of the linear harmonic oscillator, defined by the
quadratic potential
\begin{equation}\label{eq_3_2_1}
V\left(x\right)=\frac 1 2\,m \omega_0^2 x^2
\end{equation}
In the standard representation of the Schr\"odinger equation (\ref{eq_2_1}),
it is well known that the energy spectrum is discrete, and the energy
eigenvalues are
\begin{equation}\label{eq_3_2_2}
E_n=\left(n+\frac 1 2\right)\hbar\,\omega_0\qquad\qquad n=0,1,2,\dots
\end{equation}
On the contrary, in Section \ref{sec_2} we saw that for a quadratic
potential the quantum evolution in the density matrix representation is
equivalent to the classical evolution: the physical quantum trajectories
are then given by
\begin{equation}\label{eq_3_2_3}
\varphi_Q\left(x,y,t\right)=\delta\Big(x_S-x_{cl}\left(t\right)\Big)e^{
\frac{\scriptstyle i}{\scriptstyle\hbar}\,p_{cl}\left(t\right)\,x_D}
\end{equation}
where $p_{cl}$ and $x_{cl}$ belong to classical trajectories:
\begin{equation}\label{eq_3_2_4}
p_{cl}\left(t\right)=\sqrt{2m E_0}\cos\left(\omega_0 t +\phi_0\right)\qquad
\quad x_{cl}\left(t\right)=\frac 1 {\omega_0}\sqrt{\frac{2E_0} m}\sin\left(
\omega_0 t +\phi_0\right)\quad
\end{equation}
Therefore, all classical energy levels $E_0$ are also acceptable in the
quantum domain, yielding a continuous energy spectrum; this seems to be a
severe contradiction between our approach and ordinary quantum mechanics.
We will now show that this contradiction is produced by an inherent
limitation in the formalism of ordinary quantum mechanics.

In the Schr\"odinger equation for pure states, the hamiltonian
operator plays two different roles: on one hand it generates the time
evolution of the system, on the other hand it is the energy operator.
Therefore its eigenvectors are at the same time the stationary states of
the system and the energy dispersion-free states; its eigenvalues are at
the same time the natural frequencies of the system and the possible
outcomes of an energy measurement. If the natural frequencies form a
discrete set, then the energy spectrum must be discrete too.

On the contrary, in the density matrix representation we have two different
operators: the hamiltonian $H$, as expressed by (\ref{eq_2_3}) or by
(\ref{eq_2_19}), generates the time evolution of the system, and therefore
its eigenvalues are the natural frequencies; the operator $E_S=\frac{
1}{2m}\,\textrm{P}_{\!S}^2+V\!\left(\textrm{Q}_S
\right)$, which appears in the definition (\ref{eq_2_15}) of the total
energy, is the ``physical'' energy operator, and its eigenvectors are the
energy dispersion-free states. Since in the Wigner representation the
operator $E_S$ is a c-number, its eigenvalues form a continuous set and
include all the classical energy levels; this is of course a consequence of
the commutation relation $\left[Q_S,P_S\right]=0$, and is true for all
quantum systems, not only for the linear harmonic oscillator. Does this mean
that all classical energy levels are always physically realizable in the
quantum domain? No, because in general the energy eigenvectors do not belong
to physical quantum trajectories, as defined by our localization postulate
and by our quantization condition (\ref{eq_2_39}). However, in the specific
case of the linear harmonic oscillator, the answer is yes: all classical
trajectories are also quantum trajectories, and all classical energy levels
are physically realizable in the quantum domain; indeed, since the revolution
time does not depend on the energy, the condition (\ref{eq_2_39}) does not
impose any restriction on the admissible energy values.

As for the natural frequencies, they form a discrete spectrum already in the
classical statistical description of the linear harmonic oscillator: indeed,
the classical probability density $F\left(x,p,t\right)$ is periodic, i.e.
$F\left(x,p,t+T_0\right)=F\left(x,p,t\right)$ where $T_0=\frac{2\pi}
{\omega_0}$ is the revolution time of the oscillator. Therefore its Fourier
expansion contains only the integer harmonics of the fundamental frequency
$\omega_0$; these are exactly the same observable natural frequencies
provided by the standard quantum description, i.e.\ the differences between
two energy levels as defined by (\ref{eq_3_2_2}).

Thus we conclude that standard quantum mechanics predicts the correct natural
frequencies but does not in general predict the correct energy levels:
the equivalence between natural frequencies and energy levels is not a
physical law, rather it is a consequence of an inherent limitation of the
standard quantum mechanical formalism. If we switch to the density matrix
representation, and eliminate the requirement that a quantum matrix be a
product of pure states $\varphi\left(x,y\right)=\psi\left(x\right)\,\psi^*
\left(y\right)$, then we should obtain the correct quantum energy levels by
imposing our localization postulate and our quantization condition
(\ref{eq_2_39}).

The validity of our approach may be confirmed by examining the ground
state of the linear harmonic oscillator in the usual quantum mechanical
description. It is well known that this state, corresponding to the energy
level $E=\frac{1}{2}\hbar\omega_0$, is described
by the gaussian function
\begin{equation}\label{eq_3_2_5}
\psi\left(x,t\right)=\left(\frac{m\omega_0}{\pi\hbar}\right)^\frac
{\scriptstyle 1}{\scriptstyle 4} e^{-\frac{\scriptstyle m\,\omega_0}
{\scriptstyle 2\hbar}\,x^2-i\,\frac{\scriptstyle \omega_0}{\scriptstyle 2}
\,t}
\end{equation}
If we apply to this state the Moyal-Wigner transformation, we obtain again
a gaussian function
\begin{equation}\label{eq_3_2_6}
F\left(x,p,t\right)=\frac{1}{\pi\hbar}\,e^{-\frac{\scriptstyle 2}
{\scriptstyle\hbar\omega_0}\left(\frac{\scriptstyle p^2}{\scriptstyle 2m}+
\frac{\scriptstyle 1}{\scriptstyle 2} m\,\omega_0^2 x^2\right)}
\end{equation}
which is always positive and therefore may be interpreted as a joint
probability distribution in classical phase space: it reproduces the same
statistical predictions of the quantum state (\ref{eq_3_2_5}) for measurements
of position and momentum. The Wigner function (\ref{eq_3_2_6}) represents
indeed a stationary state, since the right hand side does not depend on
time; however, it is certainly not an energy eigenvector. On the contrary,
it represents an ensemble of classical particles moving with all possible
energies; the energy mean value is $\langle E\rangle\!=\!\frac{1}{2}
\,\hbar\,\omega_0$ as expected, but the standard
deviation is $\Delta E\!=\!\frac{1}{2}\,\hbar
\,\omega_0$, while we would expect it to vanish for an energy eigenvector.
This result confirms our belief that the stationary states are not
necessarily energy dispersion-free: the equivalence between stationary
states and energy eigenvectors is just a shortcoming of the usual quantum
mechanical formalism, and disappears when we switch to the density matrix
representation.

As for the standard energy eigenstates with $E>\frac{1}{2}
\hbar\omega_0$, it is well known that the corresponding Wigner functions are
negative in some phase-space regions. Therefore in our approach they have no
physical meaning, i.e.\ they do not represent neither individual particles
nor statistical ensembles.

\subsection{The delta barrier potential}\label{sec_3_3}
Let's now consider a system whose potential is given by:
\begin{equation}\label{eq_3_3_1}
V\left(x\right)=V_0\,\delta\left(x\right)
\end{equation}
A set of orthonormal eigenvectors may be chosen as follows:
\begin{equation}\label{eq_3_3_2}
f_k^1\left(x\right)=\frac1{\sqrt\pi}\sin{k x}\qquad\qquad
f_k^2\left(x\right)=\frac1{\sqrt\pi}\cos\left(k\left|x\right|-\phi_k\right)
\qquad\end{equation}
where $k>0$, while $\phi_k$ is defined by
\begin{equation}\label{eq_3_3_3}
\tan\phi_k=\frac{m V_0}{\hbar^2 k}\qquad\qquad\qquad 0<\phi_k<\frac\pi 2
\qquad\end{equation}
The energy eigenvalues associated to the above solutions are simply
\begin{equation}\label{eq_3_3_4}
E_k=\frac{\hbar^2 k^2}{2 m}\qquad\qquad
\end{equation}
as in the free particle case. A linear combination of the two states $f_k^1$
and $f_k^2$ gives the well known state
\begin{equation}\label{eq_3_3_5}
\psi\left(x\right)=\left\{\begin{array}{ll}
A_0\,e^{i k x}+A_R\,e^{-i k x}\quad&\textrm{if}\quad x<0\\[6pt]
A_T\,e^{i k x}&\textrm{if}\quad x>0
\end{array}\right.
\end{equation}
where
\begin{equation}\label{eq_3_3_6}
A_R=A_0\frac{m V_0}{i k\hbar^2-m V_0}\qquad\qquad\quad
A_T=A_0\frac{i k\hbar^2}{i k\hbar^2-m V_0}\quad
\end{equation}
Since a term $e^{\pm i k x}$ represents a plane wave with momentum $p=
\pm\hbar k$, the state (\ref{eq_3_3_5}) is usually interpreted as follows:
$A_0\,e^{i k x}$ is the incident wave, moving from the left towards the
barrier; $A_R\,e^{-i k x}$ is the reflected wave, moving away from the
barrier with negative momentum $p=-\hbar k$; $A_T\,e^{i k x}$ is the
transmitted wave, moving over the barrier with the same positive momentum
as the incident wave. However, classical mechanics predicts that no particle
can be transmitted over an infinite potential barrier; therefore, the state
(\ref{eq_3_3_5}) is an example of a non-classical feature of quantum
mechanics, the so called tunnel effect. In the rest of this section we will
examine the tunnel effect in the density matrix representation.

Our goal will be to find quantum trajectories representing transmitted and
reflected particles; these quantum trajectories will be obtained as solutions
of the simplified equations (\ref{eq_2_36})-(\ref{eq_2_38}) and will satisfy
the fundamental physical principles defined in Section \ref{sec_2}, i.e.\
unitarity, energy conservation and Ehrenfest theorem. In the case of a
reflected particle, there is a very simple solution: we already know that
the equations (\ref{eq_2_36})-(\ref{eq_2_38}) are equal to their classical
counterparts, therefore we will choose the classical solution:
\begin{equation}\label{eq_3_3_7}
\varphi^{\left(0\right)}\left(x_S,t\right)=\delta\left(x_S-v t\right)\theta
\left(-t\right)+\delta\left(x_S+v t\right)\theta\left(t\right)
\end{equation}
where $v>0$ is the speed and $\theta$ is the Heaviside step function. If we
apply (\ref{eq_2_36}) to $\varphi^{\left(0\right)}$, we obtain for the
momentum density $\mathcal{P}\left(x_S,t\right)$ the following expression:
\begin{equation}\label{eq_3_3_8}
\mathcal{P}\left(x_S,t\right)=-i\hbar\varphi^{\left(1\right)}\left(x_S,t
\right)=m v\Big(\delta\left(x_S-v t\right)\theta\left(-t\right)-\delta\left(
x_S+v t\right)\theta\left(t\right)\Big)
\end{equation}
as expected, i.e.\ the incident wave-packet has momentum $P=m v$ while the
reflected wave-packet has momentum $P=-m v$. Now we integrate (\ref{eq_2_37})
to obtain $\varphi^{\left(2\right)}$; the energy density $\mathcal{E}
\left(x_S,t\right)$ is then given by:
\begin{eqnarray}
\mathcal{E}\left(x_S,t\right)&=&-\frac{\hbar^2}{2 m}\,\varphi^{\left(2
\right)}\left(x_S,t\right)+V\left(x_S\right)\varphi^{\left(0\right)}
\nonumber\\[4pt]
&=&\frac 1 2\,m v^2\Big(\delta\left(x_S-v t\right)\theta\left(-t\right)+
\delta\left(x_S+v t\right)\theta\left(t\right)\Big)\label{eq_3_3_9}
\end{eqnarray}
The derivation of (\ref{eq_3_3_9}) is not straightforward, due to the
products of distributions appearing in the intermediate calculations.
However, it may be easily obtained by replacing (\ref{eq_3_3_7}) with the
classical solution associated to the triangular potential $V_\epsilon\left(
x\right)=\frac{V_0}\epsilon\left(1-\frac{\left|x\right|}\epsilon\right)
\theta\!\left(1-\frac{\left|x\right|}\epsilon\right)$ and then taking the
limit as $\epsilon\to 0$; of course, as $\epsilon\to 0$ the function
$V_\epsilon\left(x\right)$ approaches the delta barrier potential 
(\ref{eq_3_3_1}).

Let's now turn to the case of a transmitted particle. As a first attempt,
we choose the classical solution for a free particle:
\begin{equation}\label{eq_3_3_10}
\varphi^{\left(0\right)}\left(x_S,t\right)=\delta\left(x_S-v t\right)
\end{equation}
even if we know that it can't be the final solution, since it violates
the Ehrenfest theorem at $t=0$. From (\ref{eq_2_36}) we then obtain
\begin{equation}\label{eq_3_3_11}
\varphi^{\left(1\right)}\left(x_S,t\right)=i\frac{m v}\hbar\,\delta\left(
x_S-v t\right)
\end{equation}
which is again equal to the free particle case, since equation
(\ref{eq_2_36}) does not involve the potential $V\left(x_S\right)$. The
next step would be to integrate (\ref{eq_2_37}) and find an expression
for $\varphi^{\left(2\right)}$; this would produce the following result:
\begin{equation}\label{eq_3_3_12}
\varphi^{\left(2\right)}\left(x_S,t\right)=\frac m{\hbar^2}\left(-m v^2
\delta\left(x_S-v t\right)+V_0\,\delta\left(x_S\right)\delta\left(v t\right)
+V_0\,\delta'\left(v t\right)\theta\left(x_S\right)\right)
\end{equation}
which does not satisfy the condition (\ref{eq_2_33}) at $t=0$ and therefore
violates the Ehrenfest theorem; responsible for this violation is the step
function $\theta\left(x_S\right)$ appearing in the r.h.s. of
(\ref{eq_3_3_12}). Similar problems arise when we try to integrate
(\ref{eq_2_38}): the resulting expression for $\varphi^{\left(3\right)}
\left(x_S,t\right)$ does not satisfy condition (\ref{eq_2_34}) at $t=0$,
which means that energy conservation is violated too.

Thus we have shown by direct calculations what we already knew from physical
considerations: the classical free field solution (\ref{eq_3_3_10}) may not
represent a non-classical tunnelling particle. However, since we now know the
mathematical source of the above violations, we may find a way to circumvent
them: we simply must add countertems to (\ref{eq_3_3_10}), so that the
unacceptable terms appearing in the expressions for $\varphi^{\left(2\right)}
\left(x_S,t\right)$ and $\varphi^{\left(3\right)}\left(x_S,t\right)$ are
canceled by the new ones. The resulting expression is:
\begin{equation}\label{eq_3_3_13}
\varphi^{\left(0\right)}\left(x_S,t\right)=\delta\left(x_S-v t\right)
-\frac{g_1^{\left(4\right)}\left(x_S\right)}{g_1^{\left(4\right)}\left(0
\right)}\,\delta\left(v t\right)
+\frac{g_2^{\left(3\right)}\left(x_S\right)}{g_2^{\left(4\right)}\left(0
\right)}\,\delta'\left(v t\right)\qquad
\end{equation}
where $g_1\left(x_S\right)$ and $g_2\left(x_S\right)$ are finite
approximations of the Dirac $\delta$ distribution, i.e.\ even functions
localized around $x_S=0$; $g_1^{\left(n\right)}$ and $g_2^{\left(n\right)}$
are their n-th derivatives. From the fact that $g_1$ and $g_2$ are even we
deduce that the derivatives $g_1^{\left(n\right)}\left(0\right)$ and
$g_2^{\left(n\right)}\left(0\right)$ vanish when $n$ is odd. The expressions
for $\varphi^{\left(1\right)}\left(x_S,t\right)$ and $\varphi^{\left(2
\right)}\left(x_S,t\right)$ are then:
\begin{eqnarray}
\varphi^{\left(1\right)}\left(x_S,t\right)&=&i\frac{m v}\hbar\,\left(\delta
\left(x_S-v t\right)+\frac{g_1^{\left(3\right)}\left(x_S\right)}{g_1^{\left(
4\right)}\left(0\right)}\,\delta'\left(v t\right)-\frac{g_2''\left(x_S
\right)}{g_2^{\left(4\right)}\left(0\right)}\,\delta''\left(v t\right)
\right)\label{eq_3_3_14}\\[6pt]
\varphi^{\left(2\right)}\left(x_S,t\right)&=&-\frac{m^2 v^2}{\hbar^2}\,
\left(\delta\left(x_S-v t\right)-\frac{g_1''\left(x_S\right)}{g_1^{\left(
4\right)}\left(0\right)}\,\delta''\left(v t\right)+\frac{g_2'\left(x_S
\right)}{g_2^{\left(4\right)}\left(0\right)}\,\delta^{\left(3\right)}
\left(v t\right)\right)\qquad\label{eq_3_3_15}
\end{eqnarray}

The step function has disappeared from (\ref{eq_3_3_15}), so that
$\varphi^{\left(1\right)}\left(x_S,t\right)$ and $\varphi^{\left(2\right)}
\left(x_S,t\right)$ now satisfy both conditions (\ref{eq_2_32}) and
(\ref{eq_2_33}), required respectively by the unitarity principle and by
the Ehrenfest theorem. As for $\varphi^{\left(3\right)}\left(x_S,t\right)$,
we do not write its final expression, since it does not correspond to an
observable physical quantity; it is enough to say that it now satisfies
the condition (\ref{eq_2_34}) required by energy conservation. From
(\ref{eq_3_3_14}) and (\ref{eq_3_3_15}) we may then obtain the
momentum density $\mathcal{P}\left(x_S,t\right)=-i\hbar\varphi^{\left(1
\right)}\left(x_S,t\right)\,$ and the energy density $\mathcal{E}\left(x_S,t
\right)=-\frac{\hbar^2}{2 m}\,\varphi^{\left(2\right)}\left(x_S,t\right)+
V\left(x_S\right)\varphi^{\left(0\right)}\left(x_S,t\right)$.

We further note that the wave-packet defined by (\ref{eq_3_3_13}) is sharply
localized at $x_S=v t$ when $t\ne 0$, while it has a non-vanishing
dispersion around $x_S=0$ when $t=0$, i.e.\ when it crosses the barrier. This
temporary loss of localization was expected, since it has been already
recognized in the previous section as a necessary condition for satisfying
the Ehrenfest theorem in the case of a non-classical tunnelling particle.

From what we have seen until now, we may say that (\ref{eq_3_3_7}) and
(\ref{eq_3_3_13}) represent quantum trajectories associated respectively
to reflected and transmitted particles, since they satisfy our localization
postulate together with the three fundamental physical principles. It
remains to show that they may be extended to complete solutions of the
Schr\"odinger equation, i.e.\ that they may be expressed as linear
combinations
\begin{equation}\label{eq_3_3_16}
\varphi^{\left(0\right)}\left(x_S,t\right)=
\sum_{i,j=1}^2\int\!\!\textrm{d}k_1\int\!\!\textrm{d}k_2\ C_{ij}
\left(k_1,k_2\right)f_{k1}^i\left(x_S\right)\,f_{k2}^j\left(x_S\right)
e^{-\frac{\scriptstyle i}{\scriptstyle\hbar}\,\left(E_{k1}-E_{k2}\right)\,t}
\end{equation}
for some choice of the coefficients $C_{ij}\left(k_1,k_2\right)$. We will
not demonstrate it here; however, in Appendix A we will explicitly build
numerical solutions of the Schr\"odinger equation having a very strong
resemblance with (\ref{eq_3_3_7}) and (\ref{eq_3_3_13}).

Thus we have shown that in a typical quantum system, exhibiting the highly
non-classical behaviour known as tunnel effect, it is still possible to
find solutions which remain sharply localized around their center of mass
as $t\to\pm\infty$; these solutions represent wave-packets which are totally
reflected or totally transmitted. It is our belief that they represent the
individual physical particles; the usual quantum wave-packets, which split
into a reflected part and a transmitted part, are just a statistical
mixture of our quantum trajectories. Besides, those quantum states which
may not be written as positive superpositions of quantum trajectories must
be rejected as physically meaningless.

As an example, we will examine again the state defined by
(\ref{eq_3_3_5}). If we consider the density matrix $\varphi(x,y)=\psi\left(
x\right)\psi^*\left(y\right)$, we may easily compute the expressions for
$\varphi^{\left(0\right)}\left(x_S,t\right)$, $\mathcal{P}\left(x_S,t\right)$
and $\mathcal{E}\left(x_S,t\right)$; in the case $x_S>0$ we obtain:
\begin{eqnarray}
\varphi^{\left(0\right)}\left(x_S,t\right)&=&\left|A_T\right|^2
\label{eq_3_3_17}\\[8pt]
\mathcal{P}\left(x_S,t\right)\quad &=&\left|A_T\right|^2\hbar k
\label{eq_3_3_18}\\[4pt]
\mathcal{E}\left(x_S,t\right)\quad&=&\left|A_T\right|^2
\frac{\hbar^2 k^2}{2 m}\label{eq_3_3_19}
\end{eqnarray}
The physical interpratation of (\ref{eq_3_3_17})-(\ref{eq_3_3_19}) is
straightforward: they describe an ensemble of transmitted particles
uniformly distributed to the right of the barrier (the particle density
being $\left|A_T\right|^2$ per unit of length), moving with speed $v={\hbar
k}/m$, momentum $p=\hbar k$ and energy $E={\hbar^2 k^2}/{2 m}$. To the left
of the barrier, i.e.\ in the case $x_S<0$, we have:
\begin{eqnarray}
\varphi^{\left(0\right)}\left(x_S,t\right)&=&\left|A_0\right|^2\qquad\,\,
+\left|A_R\right|^2\qquad+2 Re\left\{A_0 A_R^* e^{2 i k x_S}\right\}
\label{eq_3_3_20}\\[8pt]
\mathcal{P}\left(x_S,t\right)\quad\, &=&\left|A_0\right|^2\hbar k\quad-
\left|A_R\right|^2\hbar k\label{eq_3_3_21}\\[4pt]
\mathcal{E}\left(x_S,t\right)\quad&=&\left|A_0\right|^2\frac{\hbar^2 k^2}
{2 m}+\left|A_R\right|^2\frac{\hbar^2 k^2}{2 m}\label{eq_3_3_22}
\end{eqnarray}
The terms involving $\left|A_0\right|^2$ and $\left|A_R\right|^2$ represent
respectively an ensemble of incoming and of reflected particles, with the
right values for momentum and energy, $\left|A_0\right|^2$ and $\left|A_R
\right|^2$ being the associated particle densities; besides, the relation
$\left|A_T\right|^2+\left|A_R\right|^2=\left|A_0\right|^2$ ensures that the
total number of particles is conserved at the two sides of the barrier.
However, the last term at the r.h.s. of (\ref{eq_3_3_20}) has no
physical explanation: it does not contribute to the momentum and energy
densities, so it seems to describe an ensemble of particles at rest; yet,
it has both positive and negative values depending on $x_S$, and therefore
it may not be interpreted as a particle density. For this reason, we are
forced to reject the state (\ref{eq_3_3_5}) as physically meaningless, since
it is not a positive superposition of quantum trajectories.

On the contrary, starting from our quantum trajectories (\ref{eq_3_3_7}) and
(\ref{eq_3_3_13}) it is very easy to build a statistical ensemble having
the same physical properties as (\ref{eq_3_3_5}), but with no strange
unphysical terms. We simply have to write:
\begin{equation}\label{eq_3_3_23}
\varphi^{\left(0\right)}\left(x_S,t\right)=
\int_{-\infty}^{+\infty}{\left(
\left|A_R\right|^2\varphi^{\left(0\right)}_R\left(x_S,t-t_0\right)+
\left|A_T\right|^2\varphi^{\left(0\right)}_T\left(x_S,t-t_0\right)
\right) v\,\textrm{d}t_0}
\end{equation}
where of course $\varphi^{\left(0\right)}_R$ is the ``reflected'' quantum
trajectory (\ref{eq_3_3_7}) while $\varphi^{\left(0\right)}_T$ is the
``transmitted'' quantum trajectory (\ref{eq_3_3_13}). This is another example
of the limitations inherent to the standard formalism of quantum mechanics,
which may be overcome by switching to the density matrix representation.

Finally, we point out that the quantum trajectories (\ref{eq_3_3_7}) and
(\ref{eq_3_3_13}) must not be considered as the only possible quantum
trajectories for the delta barrier potential: specifically, it is well known
that standard quantum wave-packets impinging onto a potential barrier may
experiment a shift in time, either a time delay or a time advance. For
instance Nakazato \cite{nak} shows, precisely in the case of the delta
barrier potential, that a gaussian wave-packet under certain conditions
splits into a reflected and a transmitted wave-packet which, in the limit
$t\to\infty$, are again approximately gaussian; both wave-packets are delayed
in time with respect to the ideal case of an instant transmission or
reflection at the barrier. Therefore, it is quite possible that quantum
trajectories may be found which are shifted in time during the interaction
with the barrier; this is again a highly non-classical feature of tunnelling
particles, since of course in the classical case the interaction time with a
delta barrier vanishes and no time shift may occur.

\section{Discussion}\label{sec_4}
In the early days of quantum mechanics, Schr\"odinger tried to interpret the
solutions of his equation as ``matter waves'', i.e.\ he thought that the
individual particles could be described by sharply localized wave-packets.
Unfortunately, this realistic interpretation of the wave-function was soon
abandoned, when it became clear that wave-packets lose their localization
with time and become indefinitely extended as $t\to\infty$; this is clearly
incompatible with the fact that particles are always detected in small
space regions. Then Born proposed his interpretation of the wave-function as
``probability waves'', and his statistical postulate, incorporated in the
standard Copenhagen interpretation, has survived until today. The theory
built upon this postulate, standard quantum mechanics, gives predictions
which are in spectacular agreement with all experiments ever performed and
was successfully extended to include relativistic phenomena, such as particle
creation/annihilation, and to describe all the relevant interactions of the
microphysical world.

Nevertheless, the history of quantum mechanics is not only a history of
experimental successes; it is also a history of never ending debates about
its foundations. Countless interpretations have been proposed, alternative
to the standard Copenhagen interpretation, but none of them has been capable
of gaining universal acceptance inside the scientific community: the
supporters of the various interpretations still debate throughout the
specialized literature, emphasizing both the merits of their favourite
theories and the flaws of the opponent interpretations. It is not my
intention here to give a detailed description of the fundamental problems
which still prevent a completely satisfactory interpretation of quantum
mechanics; I just want to focus the reader's attention about a few concepts
which represent the basis on which the present paper is built.
\addtolength{\parskip}{5pt}

\underline{Realism}. A physical system exists as an objective reality,
independently from the presence of an observer performing measurements on it;
therefore a sound physical theory should be able to describe the objective
properties of the system. On the contrary, standard quantum mechanics only
provides statistical predictions about the possible outcomes of experiments,
denying the existence of objective properties prior to measurement: the
famous Schr\"odinger cat is described by a superposition of ``dead'' state
and ``alive'' state, until somebody opens the box and the cat suddenly
collapses to either ``dead'' or ``alive''. Transposing the paradox to the
description of a single particle, we have a wave-function which may be extended
over an arbitrarily large space region; then the particle's position is
measured and the wave-function suddenly collapses to the state $\left|x_0
\right>$, where $x_0$ is the measurement's result. A much more realistic
view would be to think that immediately before the measurement the particle
is already localized at the position $x_0$, this position being an objective
property of the particle: the measurement simply brings to light this property
already possessed by the particle. I am aware that such way of thinking
would be labeled as ``na\"\i f realism'' by the majority of physicists;
nevertheless, I am still convinced that it is the only acceptable starting
point for building a sound physical theory.

\underline{Determinism}. Probabilities are a very useful tool for handling
systems whose state, for whatever reason, is not completely known: they
allow us to make calculations and to obtain interesting results even if we
are partially ignorant about the system's state. Standard quantum
mechanics has radically changed this original meaning of the probability
concept: on one hand the wave-function is assumed to represent a complete
description of an individual physical system, on the other hand the Born
postulate implies that we cannot predict with absolute certainty the result
of a measurement performed on a system, even if its wave-function is
perfectly known. Thus the concept of uncertainty becomes a fundamental
principle of the theory, and the Heisenberg uncertainty relations play a
central role in the so-called ``quantum revolution''.
\addtolength{\parskip}{-5pt}

For many years it was generally believed that a fully deterministic theory
reproducing the same statistical predictions of quantum mechanics could not
exist; Von Neumann's proof of the impossibility of ``hidden variables''
theories was considered as the last word about the subject. However, in 1952
Bohm \cite{bohm} was able to build exactly this kind of theory: in Bohmian
mechanics, the state of a quantum particle is given by its wave-function
$\psi\left(x\right)$ and by its position $Q$. The time evolution of the
wave-function follows the usual Schr\"odinger equation, while the time
evolution of the particle's position is obtained from a new equation,
satisfying the following remarkable property: if we consider at time $t=0$
a statistical ensemble of particles, all described by the same wave-function
$\psi\left(x,0\right)$, whose positions $Q$ are distributed with a probability
density $\rho\left(x,0\right)=\left|\psi\left(x,0\right)\right|^2$, then the
equality  $\rho\left(x,t\right)=\left|\psi\left(x,t\right)\right|^2$ will
remain true for all times $t>0$, thus reproducing the same probabilities of
standard quantum mechanics for measurements of position. Hence Bohmian
mechanics is at the same time realistic, the position $Q$ being considered
as an objective property of the particle, and deterministic, since the
result of a measurement performed on the system is uniquely determined by
the system's state (here the assumption is made that only those observables
which may be expressed in terms of position measurements have physical
meaning).

Of course, the existence of at least one ``hidden variables'' theory
equivalent to standard quantum mechanics is enough to disprove all claims
about the impossibility of such theories.
In the present paper, as in the previous one, I strongly support the view
that the fundamental laws of nature must be deterministic: if there is
uncertainty about the results of measurements performed on a physical system,
this simply means that the state of the system is not entirely known. For
instance, let's consider two particles impinging onto the same potential
barrier, both starting with the same initial position and momentum; if the
first particle is reflected while the second one is transmitted, this is
not a consequence of some fundamental uncertainty in the laws of nature:
on the contrary, this means that position and momentum are not a complete
description of the particle's state.

Nevertheless, there is one serious
problem that must be faced by every deterministic theory trying to reproduce
the statistical predictions of quantum mechanics: of all possible probability
distributions over the system's state, why only a few may be observed in real
experiments, while the majority of them seems to be forbidden by nature? In
the case of tunnelling, we can imagine a statistical ensemble of particles
all being reflected (or transmitted); however, the experimental evidence tells
us that the real ensembles are always formed by both reflected and transmitted
particles, and their relative weight in the mixture depends on the energy
of the ensemble in a well defined manner. In Bohmian mechanics, we could
think of an ensemble of particles having all the same initial wave-function
$\psi\left(x,0\right)$ and the same initial position $Q_0$; this ensemble
would have a dispersion-free position at all future times, thus contradicting
the quantum principle that a particle's trajectory may not be perfectly known
at all times.

To solve this problem, the supporters of ``hidden variables'' theories
usually propose to complete the deterministic equations of motion with some
kind of statistical postulate, in analogy with the classical case where the
thermodynamic equilibrium hypothesis (Gibbs postulate) allows to obtain the
laws of statistical mechanics from the deterministic laws of newtonian
mechanics. In the case of Bohmian mechanics, this statistical postulate is
the already mentioned hypothesis that $\rho\left(x,0\right)=\left|\psi\left(
x,0\right)\right|^2$ at some initial time $t=0$; in the literature, this
postulate is usually referred to as the ``quantum equilibrium hypothesis''
\cite{gold}.
\addtolength{\parskip}{5pt}

\underline{Simplicity}. The history of physics supports the belief that the
fundamental laws of nature should be as simple as possible; even if this is
not a clearly defined principle, it generally means that the fundamental
objects of a theory, together with its fundamental equations, should be
reduced to the smallest possible number. For instance, the theory of special
relativity unified the electric field and the magnetic field into a single
object, the electromagnetic field, at the same time reducing the four Maxwell
equations to a single covariant expression; besides, the elimination of
absolute space and time brought as a consequence that the laws of physics
must be the same in all inertial reference frames (Lorentz invariance).
Then came the general theory of relativity, which further simplified the
definition of reference frame by eliminating the priviliged role of inertial
frames and by establishing the equivalence between accelerated frames and
gravitational fields; as a consequence, the laws of physics must be the same
in \emph{all} reference frames, inertial or not (general covariance). Besides,
general relativity provides an elegant explanation of the gravitational forces
as effect of space-time curvature and clarifies the status of space-time,
which becomes a true physical object, not only an arena where physical events
take place; it is generally believed that the  Einstein's equations for the
gravitational field are one of the highest peaks reached by the physical
science in its search for simplicity.
\addtolength{\parskip}{-5pt}

Unfortunately, the same level of simplicity has never been reached by
quantum mechanics and quantum field theory; on the contrary, there are many
controversial questions, ranging from the interpretational problems to the
divergencies which plague quantum field theory and which still prevent the
unification with gravity. Here I just want to concentrate about one point,
which is relevant to the approach that I am proposing in the present paper:
the so called wave-particle duality. It is well known that particles manifest
wave-like behaviours in some experimental situations: for instance the
interference fringes obtained in the famous two-slit experiment appear to
be a consequence of the linear superposition of two waves, adding up where
the phase is the same and canceling each other where they have opposite
phases. At the same time, the individual particles are always detected as
(almost) point-like objects, i.e.\ each of them leaves a well localized spot
on the screen: the interference fringes may be seen only after a big number
of particles has been detected. This wave-particle duality is not explained
by standard quantum mechanics, rather it is considered as a postulate, not
requiring further explanations; embarrassing questions such as which slit did
the particle pass through are rejected as meaningless, since in standard
quantum mechanics there is no room for such an old-fashioned concept as the
trajectory of a particle.

On the contrary, Bohmian mechanics provides a clear
explanation of wave-particle duality: in the case of the two-slit experiment
the wave function passes through both slits, while the particle trajectory,
expressed by the function $Q\left(t\right)$, passes through but one of the
slits. However, the point that I want to underline here is that this
explanation does not meet the simplicity criterium defined above: instead of
unifying the two concepts (particles and waves) into one single object,
capable of manifesting both behaviours, the duality is solved by stating that
particles and waves exist as two separate entities: the number of fundamental
objects of the theory, together with its fundamental equations, is increased
with respect to standard quantum mechanics, while the predictive power of the
theory remains the same. This is probably the main reason why Bohmian
mechanics has been considered by the majority of physicists more like a
mathematical curiosity, rather than a serious alternative to standard quantum
mechanics.

As a counteraxample, I will now briefly outline how the wave-particle
duality is solved by a recently proposed theory, namely Hasselmann's ``metron
model'' \cite{hass}. In this theory, the only fundamental object is the metric
tensor $g_{LM}$ defined over a higher-(eight- or nine- )dimensional space,
where the first four dimensions define the ordinary space-time, while the last
four or five are extra-space dimensions. The only fundamental equation is the
Einstein's gravitational field equation
\begin{equation}\label{eq_4_1}
R_{LM}=0
\end{equation}
where $R_{LM}$ is the Ricci curvature tensor. The author shows that the
simple equation (\ref{eq_4_1}) has ``an extremely rich nonlinear structure
which encompasses all the principal interactions of quantum field theory and
can be used as the foundation of a unified deterministic theory of fields
and particles''. Specifically, it is postulated that soliton-type solutions
exist, named ``metrons'', consisting of ``a localized, strongly non linear
core and a set of linear far fields ... The core is the origin of the
corpuscular properties of matter, while the ... far fields give rise to the
wave-like interference phenomena''. Thus the metron model unifies gravity
with the other forces of nature and solves the wave-particle duality, and all
this is achieved starting from equation (\ref{eq_4_1}), which is even simpler
than the original Einstein's field equations: indeed the energy-momentum
tensor, which appears as an external source term in the original field
equations for gravity, is not present in the fundamental equation of the
metron model; on the contrary, the author shows that the standard
energy-momentum tensor arises from the contraction of the extra-space
components of the Riemann curvature tensor. Hasselmann's model is a striking
example of a theory which meets our simplicity criterium; besides, it has
more than one point in common with the approach that I propose in the
present paper.
\addtolength{\parskip}{5pt}

So far we have identified three important guidelines for building a sound
physical theory: realism, determinism, simplicity; now let's see how the
approach presented in this paper tries to meet these criteria. Basically,
what I propose is to revive the original Schr\"odinger's idea of the
wave-function as ``matter waves''. The main obstacle to this idea, i.e.\ the
fact that wave-packets spread out with time, is overcome by choosing as the
fundamental equation of the theory the Schr\"odinger equation for density
matrices, instead of the one for pure states on which standard quantum
mechanics is based. In Section \ref{sec_2} we have seen that in the density
matrix representation the Heisenberg uncertainty principle is not true, i.e.\
there are fields perfectly localized both in position space and in momentum
space; as a consequence, wave-packets do not necessarily spread out with time.
We have further postulated that the only physical solutions (here ``physical''
means ``representing individual particles'') are those which remain well
localized as $t\to\pm\infty$, labeled ``quantum trajectories''. In Section
\ref{sec_3} we have seen that quantum trajectories exist in three important
cases: the free particle, the linear harmonic oscillator and the delta
barrier potential; I am firmly convinced that they may be proved to exist in
all cases of physical interest.
\addtolength{\parskip}{-5pt}

Our approach is deterministic: if we know the state $\varphi\left(x,y,0\right)$
of the particle at $t=0$, then the Schr\"odinger equation (\ref{eq_2_4})
determines the state $\varphi\left(x,y,t\right)$ at all future times; besides,
the values of the observable quantities (center of mass, momentum and energy)
are uniquely determined from the state by means of equations
(\ref{eq_2_13})-(\ref{eq_2_15}). Note that the particle's position is not a
well defined concept for a wave-packet, and therefore we replace it with its
center of mass; however, since we postulate that the real wave-packets remain
sharply localized as $t\to\infty$, the wave-packet's center of mass is
practically indistinguishable from the position of a pointlike particle.

Our approach is also realistic: no ``Schr\"odinger cat'' state is allowed and
no wave-function collapse is needed; the position of the particle, i.e.\ the
center of mass of a sharply localized wave-packet, is known prior to
measurement and therefore may be considered as an objective property of the
particle. Of course, since the state $\varphi\left(x,y\right)$ depends upon
two position coordinates, our approach postulates the existence of an objective
extra-space coordinate: in addition to the ``physical'' position $x_S=(x+y)
/2$, we have the ``auxiliary'' position coordinate \mbox{$x_D=x-y$}, having observable
effects only around the point $x_D=0$. The idea of extra spacetime dimensions
has a long lasting tradition in physics, which was initiated by the original
Kaluza-Klein theories \cite{kal,kle} and has been recently revived by the
already cited Hasselmann's metron model.

As for simplicity, we have only one fundamental object, the field $\varphi
\left(x,y\right)$, and one fundamental equation, the Schr\"odinger equation
(\ref{eq_2_4}). The wave-particle duality is solved by postulating that the
physical wave-packets are sharply localized as $t\to\pm\infty$, i.e.\ when
the particles are produced and detected: this explains why the individual
particles are always experienced as (almost) pointlike objects. At the same
time, the wave-packets may spread out to some extent while interacting with
the potential $V\left(x\right)$; this temporary loss of localization is
responsible for non-classical phenomena such as tunnelling or interference,
i.e.\ for the wave-like behaviour of quantum particles. We may easily imagine
the form of a quantum trajectory in the case of the two-slit experiment: it
starts as a localized wave-packet as $t\to -\infty$, then when approaching
the potential region it splits into two separate wave-packets; each
wave-packet passes through one of the slits; then, while leaving the potential
region, the two wave-packets merge again and as $t\to +\infty$ a single
localized wave-packet is detected, producing a pointlike spot on the screen.
There is an important difference between our approach and standard QM
concerning the concept of interference: in standard QM, interference takes
place between two plane (or spherical) waves defined over wide space regions;
in our approach, interference is localized in space and time, i.e.\ it is
confined to the potential region and to the time interval when the physical
wave-packet interacts with the potential. Outside this space-time region our
wave-packets move along classical trajectories, behaving as classical free
particles; however, the interaction with the potential bends the outgoing
trajectories with respect to the classical case, producing statistical effects
which may not be explained classically, like the interference fringes in the
two-slit experiment.

The superposition principle of standard QM does not apply to our
quantum trajectories, i.e.\ a linear combination of quantum
trajectories does not in general represent an individual particle. However,
it may be accepted in its statistical meaning: a positive superposition of
quantum trajectories may be taken to represent a statistical ensemble (or
mixture) of particles. This statistical generalization, of course, is not
enough to reproduce the predictions of standard QM; we still need
some kind of statistical postulate (similar to the quantum equilibrium
hypotesis in Bohmian mechanics) to explain, for instance, the relative
frequency of transmitted and reflected particles in the case of tunnelling,
or the exact form of the interference fringes in the case of the two-slit
experiment. This remains an open issue in our approach; nevertheless, it
seems to me that the solution to this problem
should be made easier by the fact that our fundamental equation, both for
individual particles and for statistical ensembles, is exactly the same
equation which describes the time evolution of density matrices in standard
QM: we did not change the equation, we only adopted a different criterium
to decide which solutions are physically meaningful and which are not.

There is a second open issue in our approach: the definition of quantum
trajectory is slightly vague, since we did not specify exactly what we mean
by ``well localized wave-packet''; should it be a Dirac delta function, with
vanishing dispersion, or is it allowed to have a finite extension
(for instance a sharp gaussian)? Moreover, the Schr\"odinger equation
(\ref{eq_2_4}) admits solutions which do not have physical meaning, i.e.\
they are neither quantum trajectories, nor positive superpositions of quantum
trajectories; we had to introduce a specific postulate to define which are
the physical solutions, but clearly a truly fundamental theory should contain
in itself a mechanism for selecting the physical solutions and discarding
the unphysical ones. The origin of this problem is evident: it is rooted in
the linearity of the Schr\"odinger equation. It is linearity which allows,
by means of the superposition principle, to obtain finite dispersion
wave-packets starting from delta functions; it is again linearity which
allows both positive and negative superpositions, thus producing solutions
for which no physical interpretation is possible, not even at the statistical
level.

Linearity is a much celebrated feature of standard quantum mechanics: it is
often claimed that without the superposition principle the wave-like features
of elementary particles could not be explained. The mathematical apparatus
of the theory is totally based on linear objects (Hilbert spaces, matrix
operators, commutation relations, \dots); since this linear apparatus happens
to work pretty well in the non-relativistic case, it has been extended with
few modifications to describe relativistic phenomena. Even if the experimental
successes of quantum field theory may not be denied, I do not share all this
enthusiasm about linearity; on the contrary my opinion is that linearity is
responsible for some of the problems which plague quantum mechanics and
quantum field theory. At the interpretational level, it is the main origin
of confusion about the meaning of the wave-function: does it represent an
individual particle or a statistical ensemble of particles? Of course this
question may not find an answer inside a linear theory, where both quantum
trajectories (representing individual particles) and positive superpositions
of quantum trajectories (representing statistical mixtures) are
solutions of the fundamental equation of the theory! At the formal
level, I am convinced that the divergences encountered in quantum field
theory, with the consequent need of cumbersome renormalization procedures
(not always effective, as in the case of gravity), are originated by the
attempt of forcing a linear structure upon a phenomenology which is
essentially non-linear.

As a consequence, I don't believe that an attempt to refine our definition
of quantum trajectory would be much useful, if confined to the linear
Schr\"odinger equation. A much more useful effort would be trying to add
non-linear terms to the equation, with the goal of eliminating the unphysical
solutions: the only stable solutions of the new equation should be those
which represent individual particles. The premises for a non-linear
modification of the
Schr\"odinger equation have been studied by many authors (see for instance
\cite{bial,doebn,gis,miel}); these explorations have been motivated mainly
by the general observation that ``all linear equations describing the
evolution of physical systems are known to be approximations of some
nonlinear theories, with only one notable exception of the Schr\"odinger
equation'' \cite{bial}. This is undoubtedly a well founded observation;
unfortunately, our motivation is much more specific, and the existing
literature seems to be of little help in our case, especially because the
starting point of the above cited authors is usually the Schr\"odinger
equation for pure states, not the one for density matrices. Besides, we
must remember that the non-relativistic Schr\"odinger equation is just an
approximation for some relativistic equation, the Dirac equation for
fermions or the Klein-Gordon equation for bosons; it seems to me that the
search for a fundamental equation should start already at the relativistic
level, i.e.\ the non-linear modifications should be applied directly to the
Dirac or Klein-Gordon equations. An example of a non-linear modification
of the Dirac equation has already been proposed in my first paper \cite{ar},
where the non-linearity was introduced by coupling the Dirac field with the
classical electromagnetic field.

The attentive reader may have noticed that our main effort has been devoted
to the subject of open quantum trajectories: closed trajectories have been
introduced only briefly at the end of Section \ref{sec_2} and examined in
detail only for the linear harmonic oscillator (Section \ref{sec_3}). The main
conclusion of the present paper about closed quantum trajectories is the
distinction between physical energy levels and natural frequencies, two
concepts which in standard quantum mechanics are taken to coincide; in our
approach, the natural frequencies are the eigenvalues of the hamiltonian
operator $H$, defined by (\ref{eq_2_3}), while the energy levels are the
eigenvalues of the physical energy operator $E_S=\frac{1}{2m}\,\textrm{P}_{\!
S}^2+V\!\left(\textrm{Q}_S\right)$. However, since the spectrum of the
operator $E_S$ is continuum, thus including all classical energy levels,
the fact that the outcomes of energy measurements are quantized is explained
by our approach in a different way than in standard QM, namely by means of
the quantization condition (\ref{eq_2_39}). Of course this subject needs
further investigation; here I just want to point out a simple observation,
which may have some interesting physical consequences. The observation is,
trivially, that nobody has ever seen what happens inside an isolated atom:
what we see is how the atom reacts to some external excitation, for instance
the freqeuncy of the light emitted by the atom after it has been excited by
the collision with an accelerated electron; the hypotesis that the emitted
light frequencies correspond to jumps of the internal electrons between two
different energy levels, as defined by standard QM, may not be experimentally
tested. It is the only possible explanation if we confine ourselves to the
Schr\"odinger equation for pure states; however, in the density matrix
representation there may be other explanations, in terms of closed quantum
trajectories whose total energy does not necessarily coincide with the energy
levels of standard QM. In the end, the only observable effect is that the
emitted light frequencies correspond to the natural frequencies of the system
under examination, and these are the same both in standard QM and in the
density matrix representation.

The present paper has been entirely devoted to the case of a single particle;
in Appendix B we will briefly sketch a possible way of dealing with
many-particle systems in the density matrix representation. Here I just want
to point out that every realistic interpretation of a many-particle quantum
system must face the well known dilemma of quantum non-locality: the result
of a given measure performed on one particle seems to be influenced by what
kind of measurement is performed on a second particle, distant from the first.
However, this weird feature of quantum mechanics is not fully established:
there are controversies still open both at the theoretical and at the
experimental level. On one hand, there are authors \cite{sza1,sza2} who
believe that the quantum probabilities should be considered as conditional
probabilities and therefore they have no right to enter the Bell inequalities;
on the other hand there are still doubts about the experimantal violation of
Bell inequalities, due to the well known loopholes \cite{santos,kwiat}.
Therefore the impossibility of reproducing the experimental predictions of
standard QM (and the observed experimental results) by means of local realistic
theories has not yet been proved beyond doubt, and maybe never will be; my
attitude about this point is not different from the one I have taken in the
single particle case, where those quantum states which could not be expressed
as positive superpositions of quantum trajectories were rejected as physically
meaningless: in the many-particle case, I am inclined to reject as physically
meaningless those quantum states whose statistics may not be reproduced by
local realistic theories.

\newpage
\appendix
\section{Numerical quantum trajectories for the delta barrier potential}
As a first step, we normalize the Schr\"odinger equation by imposing
$m\!=\!\hbar\!=\!V_0\!=\!1$; this choice does not result in a loss of
generality, since we may get back the original Schr\"odinger equation
through a suitable rescaling of the coordinates $x$, $t$ and of the field
$\varphi$. As a second step, we must choose a finite set of eigenvectors
out of (\ref{eq_3_3_2}), since it is impossible to deal numerically with an
infinite continuum of eigenvectors; therefore we will consider only a limited
space region $-L\!<\!x\!<\!L$ and only those eigenvectors $f\left(x\right)$
which behave smoothly at the borders of such region, i.e.\ $f\left(L\right)
\!=\!f\left(-L\right)$ and $f'\left(L\right)\!=\!f'\left(-L\right)$. Our set
of eigenvectors will then be defined by:
\begin{eqnarray}
f_n\left(x\right)&=&\cos\left(k_n\left|x\right|-\phi_n\right)\qquad
\textrm{for}\ \ n\ \ \textrm{even}\label{eq_A_1}\\[6pt]
f_n\left(x\right)&=&\sin k_n x\qquad\qquad\qquad\,\textrm{for}\ \ n\ \
\textrm{odd}\label{eq_A_2}
\end{eqnarray}
For $n$ odd, $k_n$ is simply given by
\begin{equation}\label{eq_A_3}
k_n\,L=\frac{n+1}2\,\pi
\end{equation}
while for $n$ even it may be obtained by solving the coupled equations:
\begin{eqnarray}
\tan\phi_{\!n}\!&=&\frac 1{k_n}\qquad\qquad\qquad 0<\phi_n<\frac\pi 2
\label{eq_A_4}\\[6pt]k_n\;L\ &=&\phi_n+\frac n 2\,\pi\label{eq_A_5}
\end{eqnarray}
For our computations, we will choose $L=100$.

Our numerical quantum trajectories will then have the form
\begin{equation}\label{eq_A_6}
\varphi\left(x,y,t\right)=
\sum_{i,j=0}^N C_{ij}\,\varphi_{ij}\left(x,y,t\right)=
\sum_{i,j=0}^N C_{ij} f_i\left(x\right)
f_j\left(y\right) e^{i\left(\omega_j-\omega_i\right) t}\quad
\end{equation}
where $\omega_i=k_i^2/2$; our goal will be to find coefficients $C_{ij}$ such
that (\ref{eq_A_6}) provides good approximations for the simplified solutions
(\ref{eq_3_3_7}) and (\ref{eq_3_3_13}), which represent reflected and
transmitted particles. To accomplish this task, we first associate to every
quantum trajectory $\varphi\left(x_S,x_D,t\right)$ the three functions
\begin{eqnarray}
\rho\left(x_S,t\right)&=&\varphi\left(x_S,x_D,t\right)\bigg|_{x_D=0}ù
\label{eq_A_7}\\[6pt]
\textrm{\large$\wp$}
\left(x_S,t\right)&=&-i\frac\partial{\partial
x_D}\,\varphi\left(x_S,x_D,t\right)\Bigg|_{x_D=0}\label{eq_A_8}\\[6pt]
\textrm{\large$\varepsilon$}
\left(x_S,t\right)&=&\left(-\frac 1 4\frac{\partial^2}{\partial
x^2}-\frac 1 4\frac{\partial^2}{\partial y^2}+\delta\left(x_S\right)\right)
\varphi\left(x_S,x_D,t\right)\Bigg|_{x_D=0}\label{eq_A_9}
\end{eqnarray}
which represent respectively the mass
density, the momentum density and the energy density; the reader may notice
that the definition (\ref{eq_A_9}) for the energy density differs from the
usual one obtained from (\ref{eq_2_15}). The reason is that, while the total
energy is the same, the choice made in (\ref{eq_A_9}) eliminates from the
beginning the possible appearance of delta functions at $x_S=0$.

Then we define the scalar product
\begin{eqnarray}
\langle\,\varphi_1\,\varphi_2\,\rangle&=&\int_{-T}^T\!\textrm{d}t\int_{-L}^L\!\textrm{d}x_S\,
\Big(
w_0\,\rho_1\!\left(x_S,t\right)\rho_2\!\left(x_S,t\right)+\nonumber\\[6pt]
&&\qquad+\;w_1\,\textrm{\large$\wp$}_1\!\left(x_S,t\right)\textrm{\large$\wp$}_2\!\left(x_S,t\right)+
w_2\,\textrm{\large$\varepsilon$}_1\!\left(x_S,t\right)\textrm{\large$\varepsilon$}_2\!\left(x_S,t\right)
\Big)\qquad\label{eq_A_10}
\end{eqnarray}
where $w_0$, $w_1$ and $w_2$ are appropriate weights, while $\left[-T\,,T\,
\right]$ is a time interval to be defined. To obtain our numerical quantum
trajectories we will project, by means of the scalar product (\ref{eq_A_10}),
the simplified solutions (\ref{eq_3_3_7}) and (\ref{eq_3_3_13}) over the
linear space defined by (\ref{eq_A_6}). For instance, if $\varphi_R$ is the
solution associated to a reflected particle as defined by
(\ref{eq_3_3_7})-(\ref{eq_3_3_9}), the coefficients $C_{ij}$ will be
obtained by solving the system of $(N+1)^2$ linear equations
\begin{equation}\label{eq_A_11}
\sum_{i,j=0}^N C_{ij}\,\langle\,\varphi_{kl}\,\,\varphi_{ij}\,\rangle=
\langle\,\varphi_{kl}\,\,\varphi_R\,\rangle\qquad\qquad k,l=0,\dots,N
\end{equation}
which requires the inversion of a $(N+1)^2\times (N+1)^2$ square matrix. The
choice of $N$ is then dictated by the computational resources of the machine
used for performing this matrix inversion; in our case we will set $N=100$.

For both reflected and transmitted particles we will choose the speed $v=1$;
in the standard quantum mechanical treatment this is the speed for which the
transmission probability and the reflection probability are the same. Besides,
due to numerical considerations, we will not use delta functions in the
definitions of $\varphi_R$ and $\varphi_T$. On the contrary, we will use the
smooth function
\begin{equation}\label{eq_A_12}
f\left(x\right)=\frac 8{3\pi\Delta x}\cos^4\!\!\left(\frac x{\Delta x}\right)
\theta\!\left(\frac\pi 2-\frac{\left|x\right|}{\Delta x}\right)
\end{equation}
which in the limit $\Delta x\to 0$ behaves like a Dirac delta function. In
(\ref{eq_A_12}), $\theta$ is the Heaviside step function, while $\Delta x$
is a finite dispersion to be defined; for our computations we will use
$\Delta x=3$. Besides, we are now able to choose the time $T$ appearing in
the definition (\ref{eq_A_10}) of our scalar product: we will set
$T=L-\frac\pi 2\Delta x$, so that the wave-packet $f\left(x-v t\right)$
is entirely included in the space region $\left[-L\,,L\,\right]$ when
$t\in\left[-T\,,T\,\right]$.

Now we are ready to show the results of our computations. In the case of a
reflected particle, with a suitable choice of the coefficients $w_0$, $w_1$
and $w_2$, we obtain the quantum trajectory described by figure 1. In figure
1.a, we show the wave-packet representing the mass density $\rho\left(x_S
\right)$ at different times $t=\pm 80,\pm 40, 0$. In figures 1.b and 1.c,
we show the wave-packets associated to the momentum and energy densities at
the same times; the momentum density at $t=0$ may not be displayed, since it
vanishes identically for all $x_S$. Finally, in figures 1.d and 1.e we show
the mean value $x_M$ and the standard deviation $\sigma_x$ of the
wave-packet as a function of time; these two quantities have been obtained
using as weight the function $\rho^4$, for instance the mean value is given by
\begin{equation}\label{eq_A_13}
x_M=\frac{\int_{-L}^{+L} x\,\rho^4\!\left(x\right)\textrm{d}x}
{\int_{-L}^{+L}\rho^4\!\left(x\right)\textrm{d}x}
\end{equation}
This choice has been made to reduce the effect of the low amplitude noise
appearing in figure 1.a. From figure 1, it is clear that our quantum
trajectory is a very good approximation of the corresponding classical
trajectory. Specifically, the standard deviation is approxmately constant
when $t$ is different from zero, but is reduced for $t=0$, due to the
overlapping of the incoming part and the reflected part of the wave-packet;
this is qualitatively the same behaviour that may be seen in the classical
case.

Turning now to the case of a transmitted particle, we make a slight change
to the definition of our scalar product (\ref{eq_A_10}) and multiply the
integrand term by the factor
\begin{equation}\label{eq_A_14}
p\left(x_S,t\right)=1-\theta\left(\pi\Delta x-\left|x_S\right|\right)\,
\theta\!\left(\frac\pi 2\Delta x-\left|t\right|\right)
\end{equation}
that is, we eliminate from the integral the region defined by $x_S\in\left[-
\pi\Delta x\,,\pi\Delta x\,\right]$ and $t\in\left[-\frac\pi 2\Delta x\,,
\frac\pi 2\Delta x\,\right]$. The reason is that we do not want to force any
form to the wave-packet while it is tunnelling through the barrier; after all,
this is a non-classical behaviour, and we do not have any a priori knowledge
of the deformations produced on the wave-packet by this non-classical
interaction. With a suitable choice of the coefficients $w_0$, $w_1$ and
$w_2$, we then obtain the quantum trajectory described by figure 2. The time
evolution of the mean value $x_M$ (figure 2.d) is not different from the
case of a classical free particle; however, from figure 2.a we see that at
$t=0$ the wave-packet is heavily deformed, while from figure 2.e we see that
the standard deviation increases during the interaction time, as expected.

Thus we have found two complete solutions of the Schr\"odinger equation which
approximate pretty well the simplified solutions (\ref{eq_3_3_7}) and
(\ref{eq_3_3_13}). Our numerical solutions have some limitations: their
range of validity is restricted to the space-time region defined by $x_S\in
\left[-L\,,L\,\right]$ and $t\in\left[-T\,,T\,\right]$, while outside this
region the wave-packet looses abruptly its localization; besides, the
localization of our wave-packets is not so sharp even inside the region of
validity. However, it should be clear that these limitations arise
from the fact that we are working with a finite set of eigenvectors: if we
were able to perform our calculations in the limit $N\to\infty$, the
validity of our quantum trajectories could be extended to larger space-time
regions and to sharper wave-packets, thus approaching the limit $L\to\infty$,
$T\to\infty$ and $\Delta x\to 0$.

\begin{figure}
\includegraphics[scale=1.18,trim=15 0 0 0]{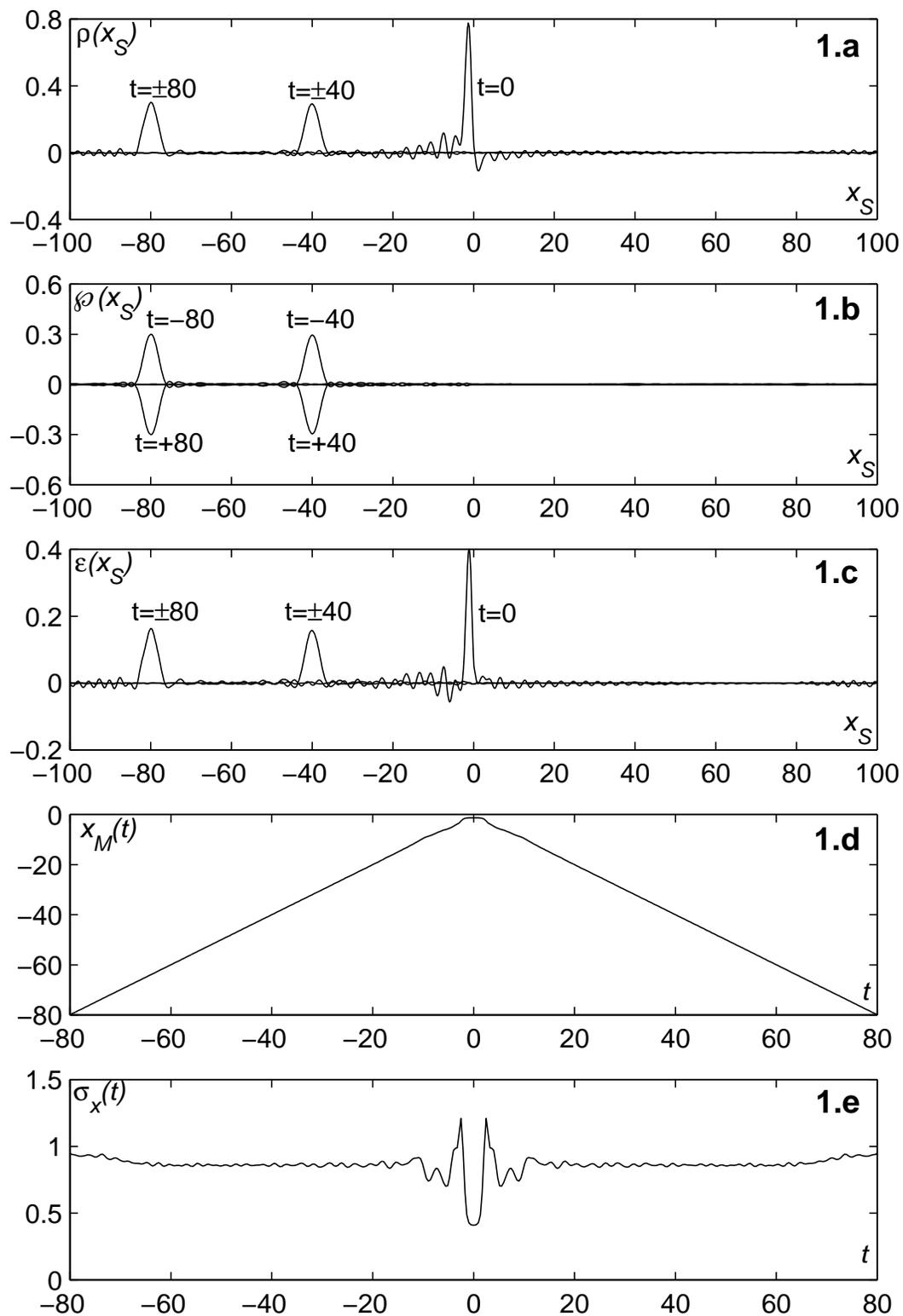}
\caption{Quantum trajectory for a reflected particle}
\label{fig_A_1}
\end{figure}

\begin{figure}
\includegraphics[scale=1.18,trim=15 0 0 0]{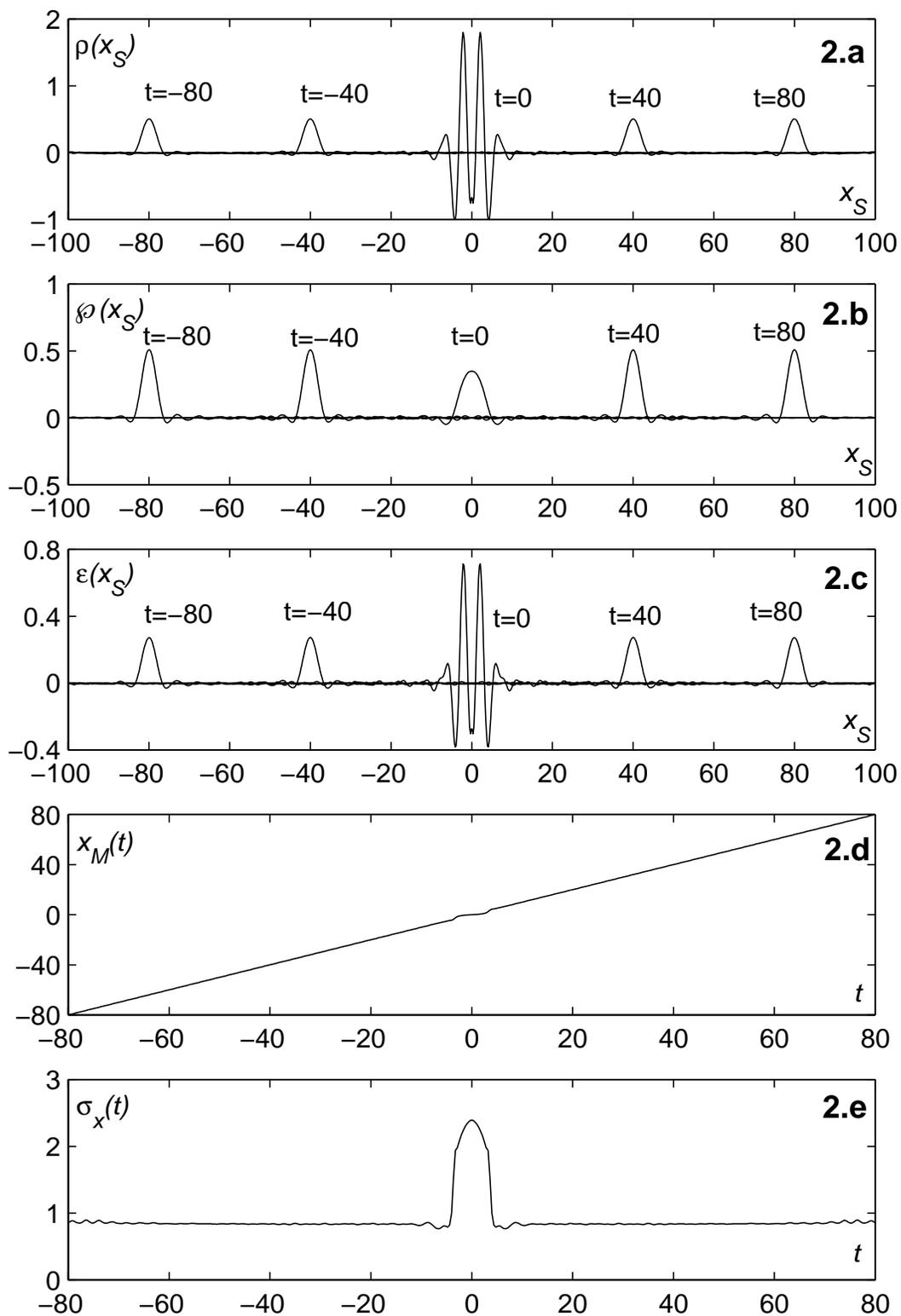}
\caption{Quantum trajectory for a transmitted particle}
\label{fig_A_2}
\end{figure}

\newpage
\section{Brief introduction to the many-particle case}
In standard QM, a system of $N$ particles
is described by a wave-function $\psi\left(x_1,\dots,x_N\right)$ and the fact
that $\psi$ is defined over a configuration space, whose dimension depends on
the particles' number, is a further obstacle to a realistic interpretation
of the wave-function; therefore this undesired feature must disappear in the
density matrix representation. To be simple, let's consider a system of two
particles moving in one space dimension under the effect of an external
potential $V\left(x_1,x_2\right)$; to avoid possible complications which
may arise in the case of identical particles, we will further suppose
that the two particles have different masses $m_1$ and $m_2$. The
Schr\"odinger equation for the density matrix $\varphi\left(x_1,x_2,y_1,y_2
\right)$ is then:
\begin{equation}\label{eq_B_1}
i\hbar\frac{\partial\varphi}{\partial t}=-\sum_{i=1}^2\frac{\hbar^2}{2m_i}
\left(\frac{\partial^2\varphi}{\partial x_i^2}-\frac{\partial^2\varphi}
{\partial y_i^2}\right)+\left[V\!\left(x_1,x_2\right)-V\!\left(y_1,y_2
\right)\right]\varphi
\end{equation}
In our approach, the solutions of equation (\ref{eq_B_1}) are taken to
represent individual physical systems instead of statistical mixtures. To
eliminate the dependence of $\varphi$ upon the configuration variables
$\left(x_1,y_1\right)$ and $\left(x_2,y_2\right)$, we would like to impose
the following separability condition:
\begin{equation}\label{eq_B_2}
\varphi\left(x_1,x_2,y_1,y_2,t\right)=\varphi_1\left(x_1,y_1,t\right)\,
\varphi_2\left(x_2,y_2,t\right)
\end{equation}
so that the two functions $\varphi_1$ and $\varphi_2$ would describe
separately the time evolution of the two particles and could be thought to
depend upon just one coordinate pair $\left(x,y\right)$, as in the single
particle case. Unfortunately, the condition (\ref{eq_B_2}) is in general too
strong, and may be satisfied only in some special cases, for instance when
$V\left(x_1,x_2\right)=V_1\left(x_1\right)+V_2\left(x_2\right)$. However,
since we know that the observable quantities depend only upon the values of
$\varphi$ in the region where $x_1\approx y_1$ and $x_2\approx y_2$, we may
impose a weaker separability condition: therefore we first define
\begin{equation}\label{eq_B_3}
x_{iS}=\frac{1}{2}\left(x_i+y_i\right)\qquad\qquad\qquad x_{iD}=x_i-y_i
\qquad\qquad i=1,2
\end{equation}
and then require that (\ref{eq_B_2}) be satisfied only far small values of
$x_{1D}$ and $x_{2D}$.

For $x_{1D}=x_{2D}=0$ we then obtain the condition
\begin{equation}\label{eq_B_4}
\varphi\Big|_{x_{1D}=x_{2D}=0}=
\rho_1\left(x_{1S},t\right)\,\rho_2\left(x_{2S},t\right)
\end{equation}
while at first order in $x_{1D}$ and $x_{2D}$ we obtain:
\begin{eqnarray}
-i\hbar\frac{\partial\varphi}{\partial x_{1D}}
\Bigg|_{x_{1D}=x_{2D}=0}&=&\mathcal{P}_1\left(x_{1S},t\right)
\,\rho_2\left(x_{2S},t\right)\label{eq_B_5}\\[6pt]
-i\hbar\frac{\partial\varphi}{\partial x_{2D}}
\Bigg|_{x_{1D}=x_{2D}=0}&=&\rho_1\left(x_{1S},t\right)\,
\mathcal{P}_2\left(x_{2S},t\right)\label{eq_B_6}
\end{eqnarray}
If we now extend the definitions (\ref{eq_2_13}) and (\ref{eq_2_14}) of center
of mass and momentum to the two-particle case, we easily obtain
\begin{eqnarray}
Q&=&Q_1+Q_2=\int{x_S\,\rho_1\left(x_{S}\right)\textrm{d}x_S}+
\int{x_S\,\rho_2\left(x_{S}\right)\textrm{d}x_S}\label{eq_B_7}\\[6pt]
P&=&P_1+P_2=\int{\mathcal{P}_1\left(x_{S}\right)\textrm{d}x_S}+
\int{\mathcal{P}_2\left(x_{S}\right)\textrm{d}x_S}\label{eq_B_8}
\end{eqnarray}
where we supposed that both $\rho_1$ and $\rho_2$ satisfy the unitarity
condition $\int{\rho_i\left(x_{S}\right)\textrm{d}x_S}=1$. As for the energy,
it is clearly impossible to define two separate energy densities, since the
potential energy depends on the position of both particles; this was true
already in the classical case.

At first order in $x_{1D}$ and $x_{2D}$, the Schr\"odinger equation
(\ref{eq_B_2}) may be decoupled in the two separate equations:
\begin{equation}\label{eq_B_9}
\frac{\partial\mathcal{P}_i}{\partial x_S}=-m\frac{\partial\rho_i}{\partial t}
\qquad\qquad\qquad i=1,2
\end{equation}
which are the natural extensions of the condition (\ref{eq_2_36}) to the
two-particle case. This decoupling of the Schr\"odinger equation confirms that
solutions satisfying to both (\ref{eq_B_4}) and (\ref{eq_B_5})-(\ref{eq_B_6})
may indeed exist.

At this point it is clear what should be the definition of quantum trajectories
in the two-particle case: we will choose those solutions which satisfy both
(\ref{eq_B_4}) and (\ref{eq_B_5})-(\ref{eq_B_6}) and which remain well
localized around their center of mass as $t\to\pm\infty$; this localization
condition applies separately to both functions $\rho_1$ and $\rho_2$.
These quantum trajectories represent individual physical systems, while
positive superpositions of quantum trajectories represent statistical
ensembles; in our approach, all other solutions are devoid of physical
meaning. Of course, to be physically acceptable, our quantum trajectories
must also satisfy the three fundamental principles, i.e.\ unitarity,
Ehrenfest theorem and energy conservation.


\newpage
\begin{thebibliography}{99}
\bibitem{bial} J. Bialynicki-Birula and J. Mycielski, Ann. Phys.
$\mathbf{100}$, 62 (1976)
\bibitem{bohm} D. Bohm, Phys. Rev. $\mathbf{85}$, 166 (1952), ibid.
$\mathbf{85}$, 180 (1952)
\bibitem{doebn} H. D. Doebner and G. A. Goldin, Phys. Rev. A $\mathbf{54}$,
3764 (1996)
\bibitem{gold} D. D\"urr, S. Goldstein and N. Zangh\`\i, Jour. Stat. Phys.
$\mathbf{67}$, 843 (1992)
\bibitem{mwi} H. Everett, Rev. Mod. Phys. $\mathbf{29}$, 454 (1957)
\bibitem{santos} M. Ferrero, S. F. Huelga and E. Santos, Phys. Rev. A
$\mathbf{51}$, 5008 (1995)
\bibitem{grw} G. C. Ghirardi, A. Rimini and T. Weber, Phys. Rev. D
$\mathbf{34}$, 470 (1986)
\bibitem{gis} N. Gisin, Phys. Lett. A $\mathbf{143}$, 1 (1990)
\bibitem{hass} K. Hasselmann, Physics Essays, $\mathbf{9}$, 311 (1996a)
\bibitem{kal} Th. Kaluza, Sitzungsber. Preuss. Akad. Wiss. Leipzig
(1921), 966
\bibitem{kle} O. Klein, Z. Phys. $\mathbf{37}$ (1926) 895
\bibitem{kwiat} P. G. Kwiat, P. H. Eberhard, A. M. Steinberg and R. Y. Chiao,
Phys. Rev. A $\mathbf{49}$, 3209 (1994)
\bibitem{miel} B. Mielnik, Commun. Math. Phys. $\mathbf{15}$, 1 (1969)
\bibitem{moy} J. Moyal, Proc. Camb. Phil. Soc. $\mathbf{45}$, 99 (1949)
\bibitem{nak} H. Nakazato, Found. Phys. $\mathbf{27}$, 1709 (1997)
\bibitem{ola} L.S.F. Olavo, Quantum Mechanics As A Classical Theory I:
Non-relativistic Theory, quant-ph/9503020
\bibitem{pearl} P. Pearle, Phys. Rev. A $\mathbf{39}$, 2277 (1989)
\bibitem{ar} A. Raiteri, A realistic interpretation of the density matrix
I: Basic concepts, quant-ph/9812011
\bibitem{schr} E. Schr\"odinger, Naturwiss. $\mathbf{14}$ (1926) 664
\bibitem{sza1} L. E. Szab\'o, Found. of Phys. Lett. $\mathbf{8}$, 421 (1995)
\bibitem{sza2} L. E. Szab\'o, Int. J. Theor. Phys. $\mathbf{34}$, 1751 (1995)
\bibitem{wig} E. Wigner, Phys. Rev. $\mathbf{40}$, 74 (1932)
\bibitem{deco1} W. H. Zurek, Phys. Rev. D $\mathbf{24}$, 1516 (1981)
\bibitem{deco2} W. H. Zurek, Phys. Rev. D $\mathbf{26}$, 1862 (1982)
\end{thebibliography}
\end{document}